\documentclass[aps,pre,preprint,superscriptaddress,nofootinbib]{revtex4-2}

\usepackage{amsmath,amssymb,amsfonts,bm}
\usepackage{graphicx}
\usepackage{dcolumn}
\usepackage{booktabs}
\usepackage{hyperref}
\usepackage{xcolor}
\usepackage{mathtools}
\usepackage{physics}
\usepackage{siunitx}
\usepackage{subcaption}
\usepackage{comment}

\newcommand{\ii}{\mathrm{i}}
\newcommand{\ee}{\mathrm{e}}

\begin{document}

\title{Interacting self-phoretic particles on a ring}

\author{Leah Anderson}
\affiliation{Université Bordeaux, CNRS, LOMA, UMR 5798, F-33400 Talence, France}
\author{David S. Dean}
\affiliation{Université Bordeaux, CNRS, LOMA, UMR 5798, F-33400 Talence, France}

\date{\today}

\begin{abstract}
We study the spontaneous motion and collective dynamics of self-phoretic particles on a ring. The particles act as localized sources of a chemical field, with particle-dependent source strengths, and respond to gradients of this field through particle-dependent phoretic mobilities. The chemical field undergoes diffusion and evaporation. For a single particle, we derive the exact threshold for spontaneous propulsion and establish the emergence of stable constant-velocity states near onset through a supercritical pitchfork bifurcation. We then consider an active source coupled to a non-emitting tracer and determine the conditions for slipping and stable phase-locked motion. For two mutually coupled sources, the stationary antipodal configuration can lose stability through either a pitchfork or a Hopf bifurcation, depending on the relative source strengths and phoretic mobilities. We determine the competing instability thresholds and compare them with direct numerical simulations of the coupled particle–field equations. Beyond onset, the resulting translating and oscillatory states provide qualitative counterparts to modes of motion observed experimentally for camphor disks in annular water channels. These results connect the onset of individual propulsion, tracer entrainment, and collective motion within a common analytically tractable framework.
\end{abstract}

\maketitle

\section{Introduction}
\label{sec:intro}
Active matter comprises systems whose constituents consume energy to
sustain dynamics far from thermodynamic equilibrium \cite{activematter}.
Among the various mechanisms that may drive active propulsion,
self-phoresis \cite{reviewphoresis} emerges from a particularly direct interplay between a particle
and its environment: the particle emits a field whose gradients,
in turn, propel it. Examples  include chemically active colloids \cite{ActiveColloidsReview} and
interfacial swimmers \cite{interfacialswimmers}. Propulsion arises as  concentration gradients can generate diffusiophoretic
slip along a particle surface \cite{anderson} or, at a fluid interface, produce
surface-tension gradients and Marangoni stresses
\cite{pimienta}. Although these mechanisms are physically quite different,
both couple particle motion to the spatial distribution of the released
chemical. Sustained propulsion therefore requires an asymmetry in
this distribution to generate a gradient. However, asymmetry need not be imposed by the
particle's structure (although in many cases it is), it can also arise due to a dynamically broken symmetry.
Indeed, even an isotropic particle can spontaneously develop an asymmetric
chemical field and select a direction of motion \cite{michelin}.
A small perturbation of the particle's motion changes the surrounding
chemical distribution; if the resulting gradients reinforce the
motion, the initially symmetric stationary state can become unstable.
Directed propulsion then emerges through spontaneous symmetry breaking. 
Spontaneous motion has been reported for symmetric camphor, benzoquinone, 
and phenanthroline disks
\cite{DiskDetcheverry,Suematsu2010Benzoquinone,Iida2010Phenanthroline}.
These systems complement camphor boats, whose structural asymmetry
selects a preferred direction
\cite{Kohira2001,Suematsu2010,Heisler2012,Camphorboat1,Camphorboat2}.
In annular channels, camphor boats exhibit homogeneous flow, congestion,
clustering, and swarming \cite{Suematsu2010,Heisler2012}, while symmetric
camphor grains support discrete and continuous density waves
\cite{Ikura2013}. Such observations motivate the question of how
spontaneous individual propulsion and chemically mediated interactions
combine to produce collective motion.

Minimal particle--field models provide a tractable setting in which to
address this question \cite{Grima,Sengupta,Taktikos,DiskDetcheverry,Ikeda2019}. Rather than prescribing a propulsion speed or
a reorientation process, as in standard active Brownian and
run-and-tumble descriptions \cite{activebrownian,runandtumble}, these
models evolve the particle positions together with the field responsible
for their motion. Diffusion and evaporation give the chemical field a finite
relaxation time, so each particle responds to a distribution shaped by
its own past trajectory and those of the other particles. Memory and
interactions therefore arise from the same field dynamics \cite{Sengupta,Taktikos, Liebchen2019}. This
framework allows the onset of propulsion and collective motion to be
studied within a common description, while isolating the effects of
geometry, source strength, and phoretic response.

In this work, we consider overdamped point particles coupled to a
chemical field, $c(x,t)$, that evolves through diffusion, evaporation, and
localised emission. Each particle responds linearly to the concentration
gradient at its position. For a single particle, this response takes
the form
\begin{equation}
    \frac{\mathrm{d}X(t)}{\mathrm{d}t}
    =
    -\lambda\,\partial_x c\bigl(X(t),t\bigr),
\end{equation}
where $\lambda$ is the phoretic mobility. We focus primarily on the
self-repulsive regime $\lambda>0$, in which the particle moves down
the concentration gradient and therefore resembles chemotactic swimmers that tend to move away from
their own chemical trail \cite{ChemotaxisSwimmers}. The model retains the feedback between particle
motion and chemical-field evolution without explicitly resolving the
fluid flow or the particle surface.

In previous work, we studied this model in a harmonic potential
\cite{HarmonicTrap} and in a finite one-dimensional channel with
reflecting boundaries \cite{1dchannel}. In both geometries, a single
particle undergoes a transition from rest to sustained oscillatory
motion above a critical mobility. The channel model was shown to be
equivalent to that of Koyano \emph{et al.} \cite{Koyano1} for studying a
camphor grain confined to a narrow water channel \cite{CamphorGrain1dbox}, allowing the onset
of motion to be characterised exactly, beyond low-order approximations.

Here, we place the particles on a one-dimensional periodic domain, or a ring.
This geometry retains a finite system size while removing the walls
and restoring continuous translational invariance. It therefore
permits uniformly translating states, corresponding to persistent
rotation around the ring \cite{Frank2014}, as well as time-dependent relative motion
between interacting particles. The periodic setting allows us to
distinguish the effects of chemical coupling from those of confining
boundaries. In particular this study is  motivated by intriguing experiments on camphor particles
in annular water channels, where a single symmetric grain can
spontaneously select persistent clockwise or anticlockwise motion
\cite{Nagayama2004}.

\begin{figure}[t]
    \centering
    \includegraphics[width=\linewidth]{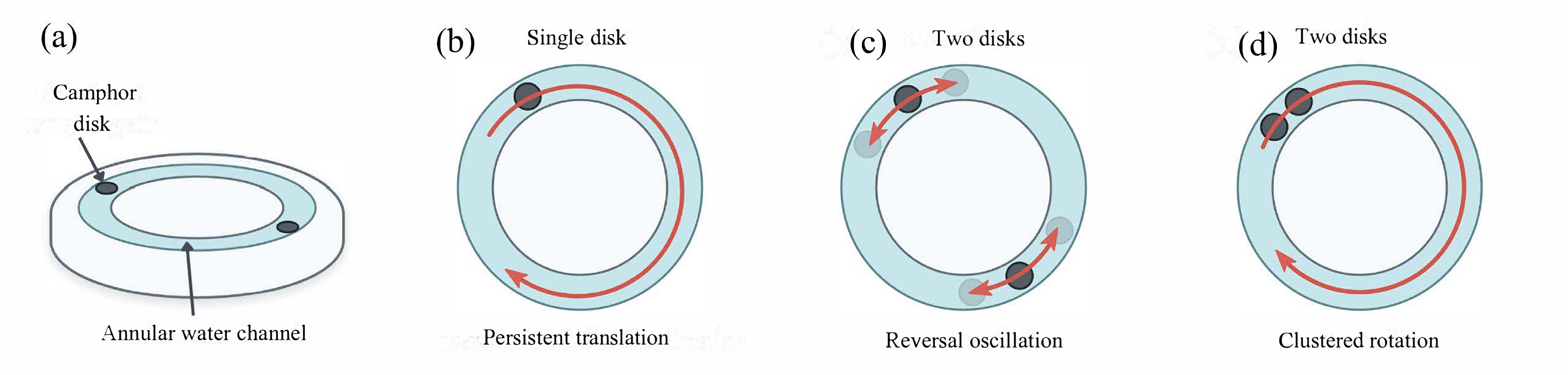}
    \caption{
        Schematic of disk-shaped camphor grains moving in an annular water channel.
        (a) Experimental geometry.
        (b) Persistent unidirectional motion of a single disk, where either the clockwise or anticlockwise direction can be selected spontaneously.
        (c) Reversal oscillation and (d) clustered rotation of two disks,
        illustrating two possible forms of collective motion observed in
        annular-channel experiments \cite{Nishi2015}.
    }
    \label{fig:camphor_schematic}
\end{figure}

The study of two identical camphor disks in an
annular water channel was carried out  by Nishi \emph{et al.} \cite{Nishi2015}.
By varying the channel diameter, they observed four types of
synchronised motion: reversal oscillation, equally spaced rotation,
clustered rotation, and stop-and-move rotation. These behaviours
were also reproduced numerically using equations of motion driven
by surface-tension differences and coupled to a reaction--diffusion
equation for the camphor concentration. The experimental geometry
and some representative motions are illustrated schematically in
Fig.~\ref{fig:camphor_schematic}. A subsequent numerical bifurcation
analysis by Yasugahira and Nagayama \cite{Yasugahira2022} identified
a broader range of dynamics, including quasiperiodic and chaotic
motion. These studies demonstrate that chemical coupling between
just two particles can support both collective rotation and
oscillatory motion, with transitions between distinct dynamical
regimes.

Our model provides a minimal description of this chemical feedback,
using overdamped point particles whose velocities respond linearly
to the local concentration gradient. This simplification permits
explicit analytical treatment of uniformly translating states and
their linear stability. Crucially, we extend the analysis to particles with different
chemical source strengths and phoretic mobilities, so that
the effects of unequal production and response can be
distinguished \cite{SotoGolestanian2014}.
Such differences are relevant to particles used in experiments, which will almost certainly exhibit variations in their
chemical loading, release rate, size, as well as their phoretic response to the gradient of the field. In addition, during the experiments it is wholly conceivable that the rates at which the particles emit vary in time as their initial chemical load is depleted. 
We use this framework to fully characterize, within our model, how particle asymmetry changes
the onset and character of collective motion. After presenting the
two-particle dynamics, we compare the translating and oscillatory
states with the modes reported by Nishi \emph{et al.}
\cite{Nishi2015}, identifying qualitative correspondences between
the minimal model and the experiments.

For a single particle, we derive the exact
activity threshold and establish the continuous emergence of
constant-velocity states through a supercritical pitchfork
bifurcation. We then consider an emitting particle coupled to a
non-emitting (or passive) tracer which nonetheless experiences a phoretic drift due to the field emitted by the emitting particle. Since the tracer does not modify the chemical field, the source retains its single-particle dynamics, while the
tracer either slips relative to it or becomes entrained at a fixed
separation. This case allows the existence and stability conditions
for phase locking to be determined explicitly. In addition this case allows a simple  analytical understanding of the behavior of the two-particle system when one of them is only weakly emitting.

When both particles emit the chemical, their motions become mutually
coupled. The separated stationary configuration is antipodal,
irrespective of differences in source strength or mobility, but
its stability depends on both. We determine the competing pitchfork
and Hopf thresholds, distinguishing the onset of collective
translation from an oscillatory instability. Direct numerical
simulations test these thresholds and characterise the long-time
motion beyond onset, connecting the linear instability to the
resulting nonlinear dynamics.

The remainder of this paper is organised as follows.
Section~\ref{sec:model_methods} introduces the governing equations
and their Fourier representation.
Section~\ref{sec:constantvelsol} derives the uniformly translating
solutions for a general $N$-particle system and develops the
linear stability framework, including the neutral mode associated
with translational invariance.
Section~\ref{sec:single_particle} examines single-particle propulsion,
its limiting cases, and the effect of periodic geometry relative
to confinement in a finite channel.
Section~\ref{sec:two_particle} treats the source--tracer system
and two mutually coupled sources, combining analytical onset
conditions with numerical investigations of the subsequent motion.
Finally, Sec.~\ref{sec:discussion_conclusion} discusses the
connection with annular-channel experiments, the limitations
of the model, and possible extensions.


\section{The periodic model}
\label{sec:model_methods}

\subsection{Governing equations}

We consider $N$ overdamped point particles on a one-dimensional
periodic domain of length $L$, equivalently a ring of circumference
$L$. The position of particle $i$ is denoted by $X_i(t)$ and is
defined modulo $L$. Each particle emits chemical at a constant rate
$s_i$ and responds to the concentration gradient generated by all
particles, including itself. The set of $(N+1)$ governing equations are given by the diffusion equation for the concentration field
\begin{equation}
    \frac{\partial c(x,t)}{\partial t}
    =
    D\frac{\partial^2 c(x,t)}{\partial x^2}
    -\mu c(x,t)
    +\sum_{j=1}^{N}s_j\delta_L\bigl(x-X_j(t)\bigr),
    \label{diffusioneqn}
\end{equation}
with the periodic boundary condition
\begin{equation}
    c(x+L,t)=c(x,t),
\end{equation}
and the overdamped Newton's equation for the particle position
\begin{equation}
    \frac{\mathrm{d}X_i(t)}{\mathrm{d}t}
    =
    -\lambda_i\,\partial_x c\bigl(X_i(t),t\bigr),
    \qquad i=1,\ldots,N,
    \label{eq:particle_real}
\end{equation}

Here, $D>0$ is the chemical diffusivity, $\mu\geq0$ is the decay
rate, and $\delta_L$ is the periodic Dirac delta function.
The coefficient $\lambda_i$ is the phoretic mobility: positive
$\lambda_i$ drives a particle down the concentration gradient,
whereas negative $\lambda_i$ drives it up the gradient. In the case where the $\lambda_i$ are positive, the particles will tend to repel each other but also be repelled from their past positions - the self-phoresis effect.
The point-source term in Eq.~\eqref{diffusioneqn} produces a discontinuity
in the spatial derivative of the concentration. To define the particle's
response at this discontinuity, we take the gradient acting at the source
to be the symmetric average of its one-sided limits,
\begin{equation}
    \partial_x c\bigl(X_i(t),t\bigr)
    \equiv
    \frac{1}{2}
    \left[
        \partial_x c\bigl(X_i(t)^+,t\bigr)
        +
        \partial_x c\bigl(X_i(t)^-,t\bigr)
    \right].
    \label{eq:symmetric_gradient_prescription}
\end{equation}
This prescription agrees with the symmetric Fourier representation
used below and specifies the self-interaction of each point
particle.

We primarily consider emitting, self-repulsive particles with
$s_i>0$ and $\lambda_i>0$. Setting $s_i=0$ gives a non-emitting
tracer that can still respond to the field produced by other
particles. The case of an attractive tracer, $\lambda_i<0$,
is briefly considered separately in Sec.~\ref{subsec:active_tracer}.
The model presented here contains no direct interparticle forces, excluded-volume
interaction, or noise; interactions arise entirely through the
shared chemical concentration field.

\subsection{Fourier representation}

The periodic geometry allows the chemical field to be decomposed
into Fourier modes, each of which relaxes independently under
diffusion and decay. We write
\begin{equation}
    c(x,t)
    =
    \sum_{n=-\infty}^{\infty}a_n(t)\phi_n(x),
\end{equation}
where the spatial eigenfunctions satisfy
\begin{equation}
    H\phi_n(x)=\epsilon_n\phi_n(x),
    \label{gg}
\end{equation}
with
\begin{equation}
    H\equiv-D\frac{\partial^2}{\partial x^2}+\mu.
\end{equation}
On a periodic domain of length $L$, the normalised eigenfunctions,
wavenumbers, and eigenvalues are
\begin{equation}
    \phi_n(x)=\frac{1}{\sqrt{L}}\ee^{\ii k_nx},
    \qquad
    k_n=\frac{2\pi n}{L},
    \qquad
    \epsilon_n=Dk_n^2+\mu.
\end{equation}
Completeness of this orthonormal basis gives the representation
\begin{equation}
    \delta_L(x-y)
    =
    \sum_{n=-\infty}^{\infty}\phi_n(x)\phi_n^*(y).
\end{equation}
Substituting these expansions into Eq.~\eqref{diffusioneqn}
and projecting onto each eigenfunction yields
\begin{equation}
    \dot a_n(t)
    =
    -\epsilon_n a_n(t)
    +
    \sum_{j=1}^{N}s_j\phi_n^*\bigl(X_j(t)\bigr).
    \label{eq:fourier_amplitudes}
\end{equation}
Thus, each mode relaxes at rate $\epsilon_n$ and is driven by
the instantaneous positions of the chemical sources. Since the concentration is real and
$\phi_{-n}(x)=\phi_n^*(x)$, its Fourier amplitudes obey
\begin{equation}
    a_{-n}(t)=a_n^*(t).
\end{equation}
It is therefore sufficient to evolve the modes with $n\geq0$
and reconstruct the field as
\begin{equation}
    c(x,t)
    =
    a_0(t)\phi_0(x)
    +
    \sum_{n=1}^{\infty}
    \left[
        a_n(t)\phi_n(x)
        +
        a_n^*(t)\phi_n^*(x)
    \right],
\end{equation}
which is manifestly real.

For a field specified at an initial time $t_0$, the solution of
Eq.~\eqref{eq:fourier_amplitudes} is
\begin{equation}
    a_n(t)
    =
    \ee^{-\epsilon_n(t-t_0)}a_n(t_0)
    +
    \sum_{j=1}^{N}s_j
    \int_{t_0}^{t}
    \ee^{-\epsilon_n(t-\tau)}
    \phi_n^*\bigl(X_j(\tau)\bigr)\,\mathrm{d}\tau.
    \label{eq:fourier_amplitudes_integral}
\end{equation}
The first term describes relaxation of the initial field, while
the integral records the chemical emitted along the particle
trajectories. Each mode therefore retains a memory of the source
positions, with relaxation time $\epsilon_n^{-1}$. For
the long-time steady-states considered below, we omit the decaying
initial-field contribution and take the source history to extend
to the remote past. For $\epsilon_n>0$, we define the retarded
inverse operator by
\begin{equation}
    \left(\epsilon_n+\frac{\mathrm{d}}{\mathrm{d}t}\right)^{-1}
    h(t)
    \equiv
    \int_{-\infty}^{t}
    \ee^{-\epsilon_n(t-\tau)}h(\tau)\,\mathrm{d}\tau.
    \label{eq:retarded_inverse_operator}
\end{equation}
The Fourier amplitudes can then be written compactly as
\begin{equation}
    a_n(t)
    =
    \left(\epsilon_n+\frac{\mathrm{d}}{\mathrm{d}t}\right)^{-1}
    \sum_{j=1}^{N}s_j\phi_n^*\bigl(X_j(t)\bigr).
\end{equation}
For $\mu>0$, this representation applies to every mode, giving
\begin{equation}
    c(x,t)
    =
    \sum_{n=-\infty}^{\infty}
    \phi_n(x)
    \left(\epsilon_n+\frac{\mathrm{d}}{\mathrm{d}t}\right)^{-1}
    \sum_{j=1}^{N}s_j\phi_n^*\bigl(X_j(t)\bigr).
    \label{eq:formal_field}
\end{equation}
Differentiating with respect to $x$ and evaluating the gradient
at each particle position yields
\begin{equation}
    \frac{\mathrm{d}X_i}{\mathrm{d}t}
    =
    -\ii\lambda_i
    \sum_{n\neq0}
    k_n\phi_n\bigl(X_i(t)\bigr)
    \left(\epsilon_n+\frac{\mathrm{d}}{\mathrm{d}t}\right)^{-1}
    \sum_{j=1}^{N}s_j\phi_n^*\bigl(X_j(t)\bigr).
    \label{eq:formal_particle_dynamics}
\end{equation}
This expression eliminates the chemical field explicitly,
leaving particle equations that depend on the past trajectories
of all sources. The spatially uniform mode $n=0$ does not contribute to the
gradient and is therefore absent from
Eq.~\eqref{eq:formal_particle_dynamics}. At each source, the symmetrically summed Fourier series converges
to the average of the one-sided gradients, consistently with
Eq.~\eqref{eq:symmetric_gradient_prescription}. Finally, we note that when $\mu=0$, the mean concentration grows continuously because
chemical is emitted but not removed. This spatially uniform
contribution has no effect on the concentration gradient and
therefore does not influence particle motion. The nonzero Fourier
modes still relax through diffusion, with rates $\epsilon_n=Dk_n^2>0$,
so Eq.~\eqref{eq:formal_particle_dynamics} remains valid in this limit.

\section{Uniformly translating states and their stability}
\label{sec:constantvelsol}

\subsection{Existence of locked solutions}

We first seek states in which all particles move with a common
velocity while maintaining fixed relative separations. These
uniformly translating, or locked, states take the form
\begin{equation}
    X_i(t)=Y_i+vt,
\end{equation}
where the $Y_i$ are constant positions in the comoving frame.
The case $v=0$ describes a stationary configuration. For this ansatz, the source contribution to each Fourier mode
has the time dependence
\begin{equation}
    \phi_n^*\bigl(X_j(t)\bigr)
    =
    \phi_n^*(Y_j)\ee^{-\ii k_nvt}.
\end{equation}
The inverse operator in
Eq.~\eqref{eq:formal_particle_dynamics} therefore acts as
multiplication by $(\epsilon_n-\ii k_nv)^{-1}$.
The time-dependent phase cancels against that from
$\phi_n(X_i(t))$, giving
\begin{equation}
    v
    =
    -\ii\lambda_i
    \sum_{j=1}^{N}s_j
    \sum_n
    \frac{
        k_n\phi_n(Y_i)\phi_n^*(Y_j)
    }{
        \epsilon_n-\ii k_nv
    },
    \qquad i=1,\ldots,N.
    \label{eq:uniform_translation_fourier}
\end{equation}
Translational invariance allows us to fix $Y_1=0$ without loss
of generality. The remaining $N$ unknowns are then the common
velocity and the $N-1$ relative particle positions. To evaluate
the Fourier sum explicitly, we factorise its denominator:
\begin{equation}
    \epsilon_n-\ii k_nv
    =
    -D\left[
        (\ii k_n)^2
        +\frac{v}{D}\ii k_n
        -\frac{\mu}{D}
    \right].
\end{equation}
For use here and in the stability analysis, we define
\begin{equation}
    r_\pm(z;v)
    =
    \frac{-v\pm\sqrt{v^2+4Dz}}{2D},
    \label{eq:rpm-z}
\end{equation}
with the square root taken on its principal branch. For real
$z>0$, this is the positive real square root. In the
locked-state equations, we abbreviate
\begin{equation}
    r_\pm\equiv r_\pm(\mu;v).
\end{equation}
We take $\mu>0$ in the following derivation; the zero-decay case
is obtained by taking the limit of the resulting expressions.
The denominator then becomes
\begin{equation}
    \epsilon_n-\ii k_nv
    =
    -D(\ii k_n-r_+)(\ii k_n-r_-),
\end{equation}
and partial-fraction decomposition yields
\begin{equation}
    \frac{\ii k_n}{\epsilon_n-\ii k_nv}
    =
    -\frac{1}{D(r_+-r_-)}
    \left[
        \frac{r_+}{\ii k_n-r_+}
        -
        \frac{r_-}{\ii k_n-r_-}
    \right].
    \label{eq:fourier_partial_fraction}
\end{equation}
Each term can now be summed using the periodic Green's function
\begin{equation}
    G_r(x,y)
    =
    \sum_n
    \frac{\phi_n(x)\phi_n^*(y)}{\ii k_n-r}.
    \label{eq:periodic_green_definition}
\end{equation}
By completeness of the Fourier basis, it satisfies
\begin{equation}
    (\partial_x-r)G_r(x,y)=\delta_L(x-y).
    \label{eq:periodic_green_equation}
\end{equation}
Away from $x=y$, the solution is exponential. Integrating across
the source gives the unit jump
\begin{equation}
    G_r(y^+,y)-G_r(y^-,y)=1.
\end{equation}
Together with periodicity, this determines
\begin{equation}
    G_r(x,y)
    =
    -\frac{\exp\bigl[r\,d_L(x,y)\bigr]}
    {\exp(rL)-1},
    \qquad x\neq y,
    \label{eq:periodic_green_solution}
\end{equation}
where
\begin{equation}
    d_L(x,y)=(x-y)\bmod L,
    \qquad 0<d_L(x,y)<L,
\end{equation}
is the directed distance from $y$ to $x$ around the domain.
Equivalently, for $x$ and $y$ in the same period,
\begin{equation}
    G_r(x,y)
    =
    \begin{cases}
    \displaystyle
    -\frac{\exp[r(x-y+L)]}{\exp(rL)-1},
    & x<y, \\[2ex]
    \displaystyle
    -\frac{\exp[r(x-y)]}{\exp(rL)-1},
    & x>y.
    \end{cases}
\end{equation}
At $x=y$, symmetric Fourier summation gives the average of
the one-sided limits:
\begin{equation}
    G_r(x,x)
    =
    -\frac{1}{2}
    \frac{\exp(rL)+1}{\exp(rL)-1}
    =
    -\frac{1}{2}\coth\left(\frac{rL}{2}\right).
    \label{eq:periodic_green_self}
\end{equation}
This value implements the symmetric prescription for the
self-interaction introduced in
Eq.~\eqref{eq:symmetric_gradient_prescription}. Substituting 
the partial-fraction decomposition and the Green's function into
Eq.~\eqref{eq:uniform_translation_fourier} gives the closed
locked-state equations
\begin{equation}
    v
    =
    \frac{\lambda_i}{D(r_+-r_-)}
    \sum_{j=1}^{N}s_j
    \left[
        r_+G_{r_+}(Y_i,Y_j)
        -
        r_-G_{r_-}(Y_i,Y_j)
    \right],
    \qquad i=1,\ldots,N.
    \label{eq:uniform_translation_closed}
\end{equation}
To express these equations compactly and prepare for the
stability analysis, we introduce the kernels
\begin{equation}
    K_z^{(q)}(x,y;v)
    =
    \frac{
        [r_+(z;v)]^qG_{r_+(z;v)}(x,y)
        -
        [r_-(z;v)]^qG_{r_-(z;v)}(x,y)
    }{
        D[r_+(z;v)-r_-(z;v)]
    },
    \qquad q=1,2.
    \label{eq:K-family}
\end{equation}
The kernel $K_z^{(1)}$ is the negative spatial derivative
of the chemical Green's function and determines the particle
drift. The kernel $K_z^{(2)}$ is the regular part of its
negative second spatial derivative and enters the linear
stability calculation; the associated delta-function
contribution is treated separately there. The locked-state equations consequently take the form
\begin{equation}
    v
    =
    \lambda_i
    \sum_{j=1}^{N}s_jK_\mu^{(1)}(Y_i,Y_j;v),
    \qquad i=1,\ldots,N.
    \label{eq:locked-equations}
\end{equation}
Here, $s_jK_\mu^{(1)}(Y_i,Y_j;v)$ is the drift generated by
source $j$ at particle $i$, per unit mobility. The terms
with $j=i$ describe self-interaction, while those with
$j\neq i$ describe chemical coupling between particles.
Solutions to \eqref{eq:locked-equations} determine the possible common velocities and
relative separations; their stability must be established
independently.

\subsection{Existence of stationary solutions}

The locked-state equations also describe stationary configurations,
obtained by setting $v=0$. For $\mu>0$, we define
\begin{equation}
    \kappa=\sqrt{\frac{\mu}{D}},
\end{equation}
so that
\begin{equation}
    r_+(\mu;0)=\kappa,
    \qquad
    r_-(\mu;0)=-\kappa.
\end{equation}
The drift kernel then reduces to
\begin{equation}
    K_\mu^{(1)}(Y_i,Y_j;0)
    =
    \frac{
        G_\kappa(Y_i,Y_j)
        +
        G_{-\kappa}(Y_i,Y_j)
    }{2D}.
    \label{eq:static_K1_green}
\end{equation}
For distinct particles, it is convenient to introduce
\begin{equation}
    d_{ij}=(Y_j-Y_i)\bmod L,
    \qquad 0<d_{ij}<L,
    \label{eq:directed_periodic_separation}
\end{equation}
the directed distance from particle $i$ to particle $j$.
With this convention, $d_L(Y_i,Y_j)=L-d_{ij}$, and hence
\begin{equation}
    G_r(Y_i,Y_j)
    =
    \frac{\ee^{r(L-d_{ij})}}{1-\ee^{rL}}.
\end{equation}
Substituting into Eq.~\eqref{eq:static_K1_green} gives
\begin{equation}
    K_\mu^{(1)}(Y_i,Y_j;0)
    =
    \frac{
        \sinh\left[\kappa(d_{ij}-L/2)\right]
    }{
        2D\sinh(\kappa L/2)
    }.
    \label{eq:static_K1_separation}
\end{equation}
Since $d_{ji}=L-d_{ij}$, the stationary drift kernel is
antisymmetric under exchange of its arguments:
\begin{equation}
    K_\mu^{(1)}(Y_j,Y_i;0)
    =
    -K_\mu^{(1)}(Y_i,Y_j;0).
    \label{eq:static_K1_antisymmetry}
\end{equation}
The symmetrically evaluated self-induced drift vanishes,
\begin{equation}
    K_\mu^{(1)}(Y_i,Y_i;0)=0,
\end{equation}
as expected for the reflection-symmetric field of a stationary
source. For $\lambda_i\neq0$, the stationary equations therefore become
\begin{equation}
    \sum_{\substack{j=1\\j\neq i}}^{N}
    s_j
    \sinh\left[\kappa(d_{ij}-L/2)\right]
    =0,
    \qquad i=1,\ldots,N.
    \label{eq:stationary_spacing_condition}
\end{equation}
Each equation requires the gradients generated by the other
particles to cancel at particle $i$. We note that the mobilities drop out
of these balance conditions: provided they are nonzero, they
do not determine the stationary separations, although they
remain relevant to stability. For equal source strengths, $s_j=s$, a solution is the uniformly
spaced configuration
\begin{equation}
    Y_i
    =
    Y_1+\frac{(i-1)L}{N}
    \pmod L,
    \qquad i=1,\ldots,N.
    \label{eq:uniformly_spaced_stationary_state}
\end{equation}
To verify this, note that the directed separations from any
particle take the values
\begin{equation}
    d_{ij}=\frac{mL}{N},
    \qquad m=1,\ldots,N-1.
\end{equation}
The contributions associated with $m$ and $N-m$ cancel:
\begin{equation}
\begin{aligned}
    &\sinh\left[
        \kappa\left(\frac{mL}{N}-\frac{L}{2}\right)
    \right]
    \\
    &\qquad+
    \sinh\left[
        \kappa\left(\frac{(N-m)L}{N}-\frac{L}{2}\right)
    \right]
    =0.
\end{aligned}
\end{equation}
When $N$ is even, the remaining term at $m=N/2$ vanishes
individually. Equal source strengths therefore admit uniform
spacing $L/N$, even when the particle mobilities differ. For unequal source strengths, uniform spacing need not satisfy
Eq.~\eqref{eq:stationary_spacing_condition}, and the separations
must be found from the coupled balance equations. An exception
is the two-particle system: for two positive source strengths
and nonzero mobilities, each balance equation contains only
one term and requires $d_{12}=d_{21}=L/2$. Thus, the separated
stationary pair is always antipodal, a result used in
Sec.~\ref{subsec:twosources}. These conditions establish the existence of stationary
configurations. To determine whether perturbations decay or
grow, we now examine the linear stability of the more general
uniformly translating states.

\subsection{Linear stability analysis of locked states}

We now examine whether a locked configuration returns to its
uniformly translating state after a small perturbation. Writing
\begin{equation}
    X_i(t)=Y_i+vt+W_i(t),
    \qquad |W_i(t)|\ll L,
    \label{eq:perturbed_positions}
\end{equation}
and expanding Eq.~\eqref{eq:formal_particle_dynamics} to first
order gives
\begin{align}
    \dot W_i(t)
    ={}&
    \lambda_i\sum_{j=1}^{N}s_j\sum_n
    k_n^2\phi_n(Y_i)\phi_n^*(Y_j)
    \Bigg[
        \frac{W_i(t)}{\epsilon_n-\ii k_nv}
        \nonumber\\
        &\hspace{2.5cm}
        -
        \ee^{\ii k_nvt}
        \left(\epsilon_n+\frac{\mathrm{d}}{\mathrm{d}t}\right)^{-1}
        \ee^{-\ii k_nvt}W_j(t)
    \Bigg]
    +f_i(t),
    \label{eq:linearized_particle_dynamics}
\end{align}
where $f_i(t)$ is an infinitesimal additive perturbation to the
particle velocity, introduced to define the linear response. The two terms in brackets have distinct physical origins.
Displacing particle $i$ changes where it samples the existing
chemical field, producing the instantaneous term proportional
to $W_i(t)$. Displacing source $j$ also changes the field itself;
the resulting perturbation diffuses and decays before affecting
particle $i$. This delayed contribution is represented by the
inverse operator acting on $W_j(t)$. The exponential factors
account for translation of the unperturbed state at velocity $v$.

Using the temporal Fourier convention
\begin{equation}
    W_j(t)
    =
    \int_{-\infty}^{\infty}
    \frac{\mathrm{d}\omega}{2\pi}\,
    \ee^{\ii\omega t}\widetilde W_j(\omega),
    \label{eq:temporal_fourier_transform}
\end{equation}
the delayed term becomes
\begin{equation}
    \ee^{\ii k_nvt}
    \left(\epsilon_n+\frac{\mathrm{d}}{\mathrm{d}t}\right)^{-1}
    \ee^{-\ii k_nvt}W_j(t)
    =
    \int_{-\infty}^{\infty}
    \frac{\mathrm{d}\omega}{2\pi}\,
    \frac{
        \ee^{\ii\omega t}\widetilde W_j(\omega)
    }{
        \epsilon_n-\ii k_nv+\ii\omega
    }.
    \label{eq:shifted_inverse_operator}
\end{equation}
Equation~\eqref{eq:linearized_particle_dynamics} therefore
takes the frequency-space form
\begin{align}
    \ii\omega\widetilde W_i
    ={}&
    \lambda_i\sum_{j=1}^{N}s_j\sum_n
    k_n^2\phi_n(Y_i)\phi_n^*(Y_j)
    \left[
        \frac{\widetilde W_i}{\epsilon_n-\ii k_nv}
        -
        \frac{\widetilde W_j}
        {\epsilon_n-\ii k_nv+\ii\omega}
    \right]
    +\widetilde f_i.
    \label{eq:linearized_frequency_equation}
\end{align}
To evaluate the spatial mode sums, we use the roots
$r_\pm(z;v)$ defined in Eq.~\eqref{eq:rpm-z}. Factorisation gives
\begin{equation}
    \frac{k_n^2}{Dk_n^2+z-\ii k_nv}
    =
    \frac{(\ii k_n)^2}
    {D[\ii k_n-r_+(z;v)][\ii k_n-r_-(z;v)]}.
\end{equation}
Polynomial division and partial-fraction decomposition then yield
\begin{equation}
\begin{aligned}
    &\frac{(\ii k_n)^2}
    {[\ii k_n-r_+(z;v)][\ii k_n-r_-(z;v)]}
    \\
    &\qquad=
    1+\frac{1}{r_+(z;v)-r_-(z;v)}
    \left[
        \frac{r_+(z;v)^2}{\ii k_n-r_+(z;v)}
        -
        \frac{r_-(z;v)^2}{\ii k_n-r_-(z;v)}
    \right].
\end{aligned}
\end{equation}
Using the periodic Green's function and completeness of the
Fourier basis, we obtain
\begin{equation}
\begin{aligned}
    \sum_n
    \frac{
        k_n^2\phi_n(x)\phi_n^*(y)
    }{
        Dk_n^2+z-\ii k_nv
    }
    &=
    \frac{
        r_+(z;v)^2G_{r_+(z;v)}(x,y)
        -
        r_-(z;v)^2G_{r_-(z;v)}(x,y)
    }{
        D[r_+(z;v)-r_-(z;v)]
    }
    +\frac{1}{D}\delta_L(x-y)
    \\
    &=
    K_z^{(2)}(x,y;v)+\frac{1}{D}\delta_L(x-y).
\end{aligned}
\label{eq:k2_mode_sum}
\end{equation}
We note that the delta-function term is independent of $z$. Its net
contribution to Eq.~\eqref{eq:linearized_frequency_equation}
is therefore formally
\begin{equation}
    \frac{\lambda_i}{D}
    \sum_{j=1}^{N}
    s_j\delta_L(Y_i-Y_j)
    \left(\widetilde W_i-\widetilde W_j\right).
    \label{eq:delta-cancellation}
\end{equation}
For a configuration of distinct particles, the terms with
$j\neq i$ vanish. For $j=i$, the term in parentheses is zero and so the two contributions
cancel mode by mode before the infinite sum is taken.
Consequently, no singular self-term remains, and the
linearised equations reduce to
\begin{equation}
    \ii\omega\widetilde W_i
    =
    \lambda_i\sum_{j=1}^{N}s_j
    \left[
        K_\mu^{(2)}(Y_i,Y_j;v)\widetilde W_i
        -
        K_{\mu+\ii\omega}^{(2)}(Y_i,Y_j;v)
        \widetilde W_j
    \right]
    +\widetilde f_i.
    \label{eq:linearized_frequency_kernel}
\end{equation}
Collecting the perturbations gives the matrix equation
\begin{equation}
    \sum_{j=1}^{N}
    \widetilde S_{ij}(\omega)\widetilde W_j(\omega)
    =
    \widetilde f_i(\omega),
    \label{eq:linearized_matrix_equation}
\end{equation}
where the inverse response matrix has elements
\begin{equation}
\begin{aligned}
    \bigl[\widetilde{\boldsymbol S}(\omega)\bigr]_{ij}
    \equiv \widetilde S_{ij}(\omega)
    ={}&
    \ii\omega\delta_{ij}
    -
    \lambda_i\delta_{ij}
    \sum_{\ell=1}^{N}
    s_\ell K_\mu^{(2)}(Y_i,Y_\ell;v)
    \\
    &+
    \lambda_i s_j
    K_{\mu+\ii\omega}^{(2)}(Y_i,Y_j;v).
\end{aligned}
\label{eq:inverse_response_matrix}
\end{equation}
The response matrix is the inverse of
$\widetilde{\boldsymbol S}(\omega)$,
\begin{equation}
    \widetilde{\boldsymbol R}(\omega)
    =
    \widetilde{\boldsymbol S}^{-1}(\omega),
\end{equation}
so that
\begin{equation}
    \widetilde{\boldsymbol W}(\omega)
    =
    \widetilde{\boldsymbol R}(\omega)
    \widetilde{\boldsymbol f}(\omega).
    \label{eq:linear_response_solution}
\end{equation}
For a stable state, the causal time-domain response is obtained
using the Fourier convention introduced above:
\begin{equation}
    R_{ij}(t)
    =
    \int_{-\infty}^{\infty}
    \frac{\mathrm{d}\omega}{2\pi}\,
    \ee^{\ii\omega t}\widetilde R_{ij}(\omega),
\end{equation}
with $R_{ij}(t)=0$ for $t<0$ and the causal prescription understood
for poles on the real axis. The poles of the response identify the linear modes. With our
Fourier convention, a mode evolves as
\begin{equation}
    \ee^{\ii\omega t}
    =
    \ee^{\ii\operatorname{Re}(\omega)t}
    \ee^{-\operatorname{Im}(\omega)t}.
\end{equation}
Modes with $\operatorname{Im}\omega>0$ therefore decay, whereas
those with $\operatorname{Im}\omega<0$ grow. A growing mode is
thus identified by a zero of
$\det\widetilde{\boldsymbol S}(\omega)$ in the lower half of
the complex $\omega$-plane.

It is important to note that one neutral mode is present for every locked configuration,
irrespective of its stability. Continuous translational
invariance implies that shifting the entire particle--field
configuration by the same amount produces an equivalent
solution. The associated perturbation has
\begin{equation}
    \widetilde W_1
    =
    \widetilde W_2
    =
    \cdots
    =
    \widetilde W_N
\end{equation}
and is proportional to
\begin{equation}
    \boldsymbol e_0=(1,1,\ldots,1)^{\mathsf T}.
\end{equation}
This neutral translational Goldstone mode occurs at $\omega=0$.
Indeed,
\begin{equation}
    K_{\mu+\ii\omega}^{(2)}(x,y;v)\big|_{\omega=0}
    =
    K_\mu^{(2)}(x,y;v),
\end{equation}
so the sum of the entries in row $i$ of the inverse response
matrix is
\begin{equation}
\begin{aligned}
    \sum_{j=1}^{N}\widetilde S_{ij}(0)
    ={}&
    -\lambda_i
    \sum_{\ell=1}^{N}
    s_\ell K_\mu^{(2)}(Y_i,Y_\ell;v)
    \\
    &+
    \lambda_i
    \sum_{j=1}^{N}
    s_j K_\mu^{(2)}(Y_i,Y_j;v)
    =0.
\end{aligned}
\end{equation}
Consequently,
\begin{equation}
    \widetilde{\boldsymbol S}(0)\boldsymbol e_0=0.
\end{equation}
This neutral mode changes only the arbitrary origin of the configuration,
leaving all relative positions unchanged. Stability is therefore
understood modulo this uniform translation. Assuming that the translational zero is simple away from a
bifurcation, we remove its factor $\ii\omega$ from the
characteristic determinant. An additional zero-frequency
mode appears when
\begin{equation}
    \lim_{\omega\to0}
    \frac{\det\widetilde{\boldsymbol S}(\omega)}
    {\ii\omega}
    =0.
    \label{eq:pitch}
\end{equation}
This is the condition for a stationary instability. In the
reflection-symmetric cases considered below, the emergence of
oppositely translating branches identifies this instability
as a pitchfork bifurcation.

An oscillatory instability instead occurs when a pair of
nonzero-frequency modes reaches the real axis:
\begin{equation}
    \det\widetilde{\boldsymbol S}(\omega_c)=0,
    \qquad
    \omega_c\in\mathbb{R}\setminus\{0\}.
    \label{eq:hopf-condition}
\end{equation}
This condition identifies a candidate Hopf threshold, with angular
frequency $|\omega_c|$ at onset. A loss of linear stability occurs
when the corresponding pair crosses from the upper to the lower
half of the complex $\omega$-plane as the coupling is increased. These zero- and finite-frequency
conditions provide the basis for the single- and two-particle
stability analyses that follow.

\section{The single-particle system}
\label{sec:single_particle}

\subsection{Constant-velocity states}

We first consider a single particle, writing $\lambda_1=\lambda$
and $s_1=s$. In this case, the uniformly translating solutions
are determined entirely by the particle's interaction with its
own chemical field. The general locked-state equation reduces to
\begin{equation}
    v=f(v)
    =\lambda sK_\mu^{(1)}(0,0;v)
    =
    \frac{\lambda s}{D(r_+-r_-)}
    \left[
        r_+G_{r_+}(0,0)-r_-G_{r_-}(0,0)
    \right],
    \label{eq:single-fixed-point}
\end{equation}
where $r_\pm=r_\pm(\mu;v)$. Substituting the self-interaction
values of the periodic Green functions gives
\begin{equation}
    v=f(v)
    =
    -\frac{\lambda s}{2D(r_+-r_-)}
    \left[
        r_+\coth\left(\frac{r_+L}{2}\right)
        -
        r_-\coth\left(\frac{r_-L}{2}\right)
    \right].
    \label{eq:single-closed}
\end{equation}

Although Eq.~\eqref{eq:single-closed} provides a compact expression
for the velocity, the Fourier representation makes the existence
and multiplicity of solutions more transparent:
\begin{equation}
    f(v)
    =
    -\frac{\lambda s}{L}
    \sum_{m\neq0}
    \frac{\ii k_m}{\epsilon_m-\ii k_mv}.
\end{equation}
Pairing positive and negative modes yields the real
self-consistency condition
\begin{equation}
    v=f(v)
    =
    \frac{2\lambda s\,v}{L}
    \sum_{m=1}^{\infty}
    \frac{k_m^2}{\epsilon_m^2+k_m^2v^2}.
    \label{eq:f-paired}
\end{equation}
Thus $f$ is odd, as required by reflection symmetry, and the
stationary state $v=0$ always exists. To determine when moving states are possible, we divide
Eq.~\eqref{eq:f-paired} by $v\neq0$ and define
\begin{equation}
    g(v)
    \equiv
    \frac{f(v)}{v}
    =
    \frac{2\lambda s}{L}
    \sum_{m=1}^{\infty}
    \frac{k_m^2}{\epsilon_m^2+k_m^2v^2}.
    \label{eq:single-gv}
\end{equation}
For $\lambda s>0$, this function is strictly decreasing
on $v>0$, since
\begin{equation}
    g'(v)
    =
    -\frac{4\lambda s\,v}{L}
    \sum_{m=1}^{\infty}
    \frac{k_m^4}
    {\left(\epsilon_m^2+k_m^2v^2\right)^2}
    <0.
    \label{eq:single-gv-derivative}
\end{equation}
Although $g(v)=f(v)/v$ is initially defined only for $v\neq0$,
it has the finite limit
\begin{equation}
    \lim_{v\to0}g(v)
    =
    \lim_{v\to0}\frac{f(v)-f(0)}{v}
    =
    f'(0).
\end{equation}
For positive velocities, $g(v)$ decreases continuously from
this value towards zero as $v\to\infty$. A moving state requires
$v=f(v)$, or equivalently $g(v)=1$. Thus, if $f'(0)\leq1$,
there is no positive-velocity solution. If $f'(0)>1$, the
function crosses unity exactly once, giving a unique positive
velocity $v_\star$. Reflection symmetry then gives a second
solution, $-v_\star$, while the stationary solution $v=0$
remains present. The condition $f'(0)=1$ therefore marks the threshold for the
existence of moving states. We next evaluate this threshold
explicitly and determine how the velocity grows close to onset.

\subsection{Activity threshold and onset of motion}

We evaluate the condition $f'(0)=1$ in terms of the dimensionless
parameters introduced in \eqref{eq:numerical_dimensionless_parameters}, which are $$ \mu^*=\frac{\mu L^2}{D},
\qquad
\lambda^*=\frac{\lambda sL}{D^2},
\qquad
\widehat v=\frac{Lv}{2D}.$$ This gives
\begin{equation}
    1=f'(0)
    =
    \frac{\lambda_c^*}{4}
    \frac{
        \sinh\sqrt{\mu^*}-\sqrt{\mu^*}
    }{
        \sqrt{\mu^*}\left(\cosh\sqrt{\mu^*}-1\right)
    }.
    \label{criteqndimless}
\end{equation}
This relation defines the critical activity $\lambda_c^*$
at each value of $\mu^*$, shown by the green curve in
Fig.~\ref{fig:critical-line-comparison}. To determine how the moving states emerge from rest, we expand
Eq.~\eqref{eq:f-paired} about $v=0$:
\begin{equation}
    f(v)
    =
    \frac{2\lambda s}{L}
    \left[
        v\sum_{m=1}^{\infty}
        \frac{k_m^2}{\epsilon_m^2}
        -
        v^3\sum_{m=1}^{\infty}
        \frac{k_m^4}{\epsilon_m^4}
    \right]
    +O(v^5).
    \label{eq:single_particle_small_v_expansion}
\end{equation}
Equivalently,
\begin{equation}
    f(v)
    =
    f'(0)v
    +
    \frac{1}{6}f'''(0)v^3
    +
    O(v^5).
\end{equation}
Substitution into the self-consistency condition $v=f(v)$
then yields
\begin{equation}
    0
    =
    \bigl[f'(0)-1\bigr]v
    +
    \frac{1}{6}f'''(0)v^3
    +
    O(v^5),
    \label{eq:single_particle_amplitude_equation}
\end{equation}
where
\begin{equation}
    f'(0)
    =
    \frac{2\lambda s}{L}
    \sum_{m=1}^{\infty}
    \frac{k_m^2}{(Dk_m^2+\mu)^2},
    \label{eq:f-prime-zero}
\end{equation}
and
\begin{equation}
    f'''(0)
    =
    -\frac{12\lambda s}{L}
    \sum_{m=1}^{\infty}
    \frac{k_m^4}{(Dk_m^2+\mu)^4}.
    \label{eq:f-third}
\end{equation}
For $\lambda s>0$, all summands are positive, so
$f'(0)>0$ and $f'''(0)<0$. The negative cubic coefficient
allows nonzero solutions to emerge on the side
$f'(0)>1$. To obtain their leading-order velocities, we write
\begin{equation}
    \delta\equiv f'(0)-1,
    \qquad 0<\delta\ll1.
\end{equation}
Dividing Eq.~\eqref{eq:single_particle_amplitude_equation}
by $v\neq0$ and rearranging gives
\begin{equation}
    v^2
    =
    \frac{6\delta}{-f'''(0)}
    +
    O(\delta^2),
\end{equation}
and hence
\begin{equation}
    v
    =
    \pm
    \sqrt{
        \frac{6\bigl[f'(0)-1\bigr]}{-f'''(0)}
    }
    +
    O\!\left(\bigl[f'(0)-1\bigr]^{3/2}\right).
    \label{eq:single_particle_velocity_near_threshold}
\end{equation}
The two moving branches therefore emerge continuously from
rest with the square-root scaling characteristic of a
supercritical pitchfork. The two signs correspond to the equivalent directions of
motion permitted by reflection symmetry. At fixed $\mu^*$, $f'(0)$ depends linearly on $\lambda^*$
and equals unity at $\lambda^*=\lambda_c^*$. Hence
\begin{equation}
    \delta=f'(0)-1
    =\frac{\lambda^*-\lambda_c^*}{\lambda_c^*},
\end{equation}
so $\delta$ measures the relative distance above the
activity threshold. Away from onset, the velocity is obtained by numerically
solving the full self-consistency condition,
Eq.~\eqref{eq:single-closed}. Having established the existence
and near-threshold behaviour of these branches, we now
examine their stability.

\subsection{Stability of the single-particle states}

For a single particle, the inverse response function is
\begin{equation}
    \widetilde S(\omega;v)
    =
    \ii\omega
    -
    \lambda s
    \left[
        K_\mu^{(2)}(0,0;v)
        -
        K_{\mu+\ii\omega}^{(2)}(0,0;v)
    \right].
    \label{eq:single-response}
\end{equation}
Using the Fourier representation of the kernels introduced in Eq.~\eqref{eq:k2_mode_sum}:
\begin{equation}
    K_z^{(2)}(0,0;v)= - \frac{1}{D}\delta_L(0) +   \sum_n
    \frac{
        k_n^2\phi_n(0)\phi_n^*(0)
    }{
        Dk_n^2+z-\ii k_nv
    },
\end{equation}
we find the reduced
characteristic function takes the form
\begin{equation}
    \frac{\widetilde S(\omega;v)}{\ii\omega}
    =
    1-
    \frac{\lambda s}{L}
    \sum_{n\neq0}
    \frac{k_n^2}{
        (\epsilon_n-\ii k_nv)
        (\epsilon_n+\ii\omega-\ii k_nv)
    }.
    \label{eq:single-reduced-response}
\end{equation}
Its zero-frequency limit is
\begin{equation}
    \lim_{\omega\to0}
    \frac{\widetilde S(\omega;v)}{\ii\omega}
    =
    1-
    \frac{\lambda s}{L}
    \sum_{n\neq0}
    \frac{k_n^2}{(\epsilon_n-\ii k_nv)^2}
    =
    1-f'(v).
    \label{eq:single-zero-frequency-condition}
\end{equation}
The general zero-frequency criterion therefore reduces to
$f'(v)=1$. On the stationary branch, this is precisely
the condition $f'(0)=1$ obtained above for the emergence
of moving states.

\begin{figure}[tb]
    \centering
    \includegraphics[width=\textwidth]{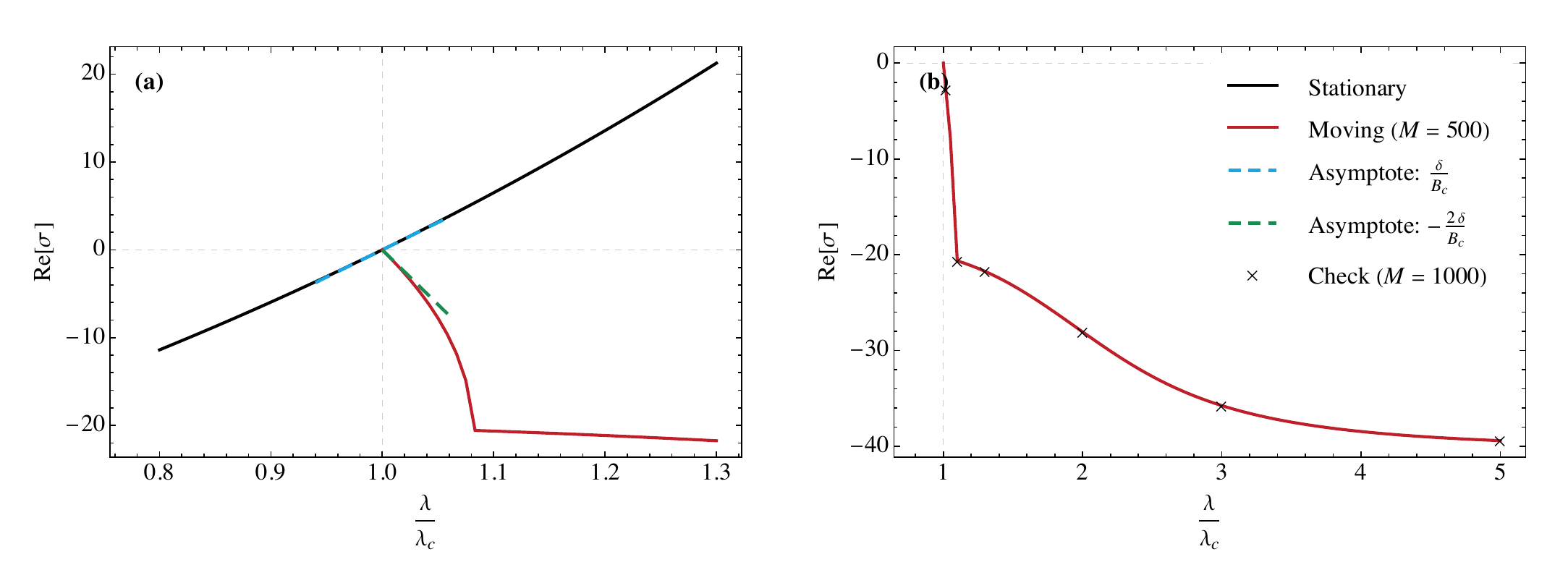}
    \caption{
    Linear stability of the single-particle states for
    $D=L=\mu=s=1$.
    (a) Leading growth rates near onset for the stationary
    (black) and moving (red) branches.
    Dashed curves show the predictions
    $\sigma_{\mathrm{st}}\simeq\delta/B_c$ (cyan) and
    $\sigma_{\mathrm{mov}}\simeq-2\delta/B_c$ (green),
    where $\delta=\lambda/\lambda_c-1$.
    (b) Leading growth rate of the moving branch over
    a wider activity range, remaining negative at all
    sampled points above onset.
    Solid curves use $M=500$; crosses show results
    for $M=1000$ at the same activities.
    The reference threshold $\lambda_c$ and the
    asymptotic coefficients in panel (a) are evaluated
    with $M=500$.
    }
    \label{fig:single_particle_stability}
\end{figure}

To determine the direction of the stability change, we
introduce $\sigma=\ii\omega$, so that perturbations evolve
as $\ee^{\sigma t}$. Expanding
Eq.~\eqref{eq:single-reduced-response} near $\sigma=0$ gives
\begin{equation}
    \frac{\widetilde S(\sigma;v)}{\sigma}
    =
    1-f'(v)+B(v)\sigma+O(\sigma^2),
    \label{eq:single-response-factored}
\end{equation}
where 
\begin{equation}
    B(v)
    =
    \frac{\lambda s}{L}
    \sum_{n\neq0}
    \frac{k_n^2}{(\epsilon_n-\ii k_nv)^3}.
    \label{eq:single-B-definition}
\end{equation}
The response pole close to the origin consequently satisfies
\begin{equation}
    \sigma
    =
    \frac{f'(v)-1}{B(v)}
    +
    O\!\left([f'(v)-1]^2\right).
    \label{eq:single-nontranslational-pole}
\end{equation}
At the activity threshold,
\begin{equation}
    B_c
    \equiv B(0)\big|_{\lambda=\lambda_c}
    =
    \frac{\lambda_c s}{L}
    \sum_{n\neq0}
    \frac{k_n^2}{\epsilon_n^3}
    >0.
    \label{eq:single-B-critical}
\end{equation}
On the stationary branch, using $\delta=f'(0)-1$ gives
\begin{equation}
    \sigma_{\mathrm{st}}
    =
    \frac{\delta}{B_c}
    +
    O(\delta^2).
    \label{eq:stationary-branch-pole}
\end{equation}
This mode therefore decays below threshold and grows above it:
the stationary state becomes unstable precisely as the moving branches
appear. For the moving states, the near-threshold expansion gives
\begin{equation}
\begin{aligned}
    f'(\pm v_\star)-1
    &=
    \delta+\frac{1}{2}f'''(0)v_\star^2+O(v_\star^4)
    \\
    &=
    -2\delta+O(\delta^2),
\end{aligned}
\label{eq:fprime-moving-branch}
\end{equation}
where we have used
$v_\star^2=-6\delta/f'''(0)+O(\delta^2)$.
Together with $B(\pm v_\star)=B_c+O(\delta)$,
Eq.~\eqref{eq:single-nontranslational-pole} therefore yields
\begin{equation}
    \sigma_{\mathrm{mov}}
    =
    -\frac{2\delta}{B_c}
    +
    O(\delta^2).
    \label{eq:moving-branch-pole}
\end{equation}
Since $B_c>0$ and $\delta>0$, the critical mode is damped
on both moving branches sufficiently close to onset. To examine stability beyond the near-threshold expansion,
we truncate the Fourier representation to $0<|n|\leq M$
and numerically diagonalise the Jacobian of the coupled
particle--field equations in the particle frame.
This gives the full eigenvalue spectrum of the truncated
system. Figure~\ref{fig:single_particle_stability}(a)
shows that the leading growth rates agree with the
analytical predictions close to onset: the stationary
state becomes unstable, while perturbations to the
moving states decay.

Farther above threshold, two real eigenvalues coalesce
and form a complex-conjugate pair, producing the change
in slope of the moving-branch curve near
$\lambda/\lambda_c\simeq1.08$.
Their real parts remain negative, indicating a crossover
to damped oscillatory relaxation towards steady translation.
The leading growth rate remains negative throughout
the sampled range $1<\lambda/\lambda_c\leq5$
[Fig.~\ref{fig:single_particle_stability}(b)].
The close agreement between the results for $M=500$
and $M=1000$ supports convergence of the leading growth
rate at the sampled activities and provides numerical
evidence for the stability of the moving states over
this range.

\subsection{Limiting cases and consistency checks}

The single-particle self-consistency condition simplifies
in two useful limits, which clarify the roles of periodicity
and chemical evaporation.

\subsubsection{Infinite-domain limit}

We first take $L\to\infty$ at fixed $\mu>0$.
Since $r_+>0$ and $r_-<0$,
\begin{equation}
    \coth\left(\frac{r_+L}{2}\right)\to1,
    \qquad
    \coth\left(\frac{r_-L}{2}\right)\to-1.
\end{equation}
Equation~\eqref{eq:single-closed} therefore becomes
\begin{equation}
\begin{aligned}
    v
    &=
    -\frac{\lambda s}{2D(r_+-r_-)}
    (r_++r_-)
    \\
    &=
    \frac{\lambda s\,v}
    {2D\sqrt{v^2+4\mu D}}.
\end{aligned}
\label{eq:infinite_domain_fixed_point}
\end{equation}
Besides the stationary solution $v=0$, the nonzero
solutions are
\begin{equation}
    v
    =
    \pm
    \sqrt{
        \left(\frac{\lambda s}{2D}\right)^2
        -4\mu D
    },
    \label{eq:infinite_domain_velocity}
\end{equation}
and exist for
\begin{equation}
    \lambda s>4\sqrt{\mu D^3}.
    \label{eq:infinite_domain_threshold}
\end{equation}
This recovers the unconfined-particle result of
Ref.~\cite{1dchannel}, where the source strength was
set to $s=1$.
The persistence of a finite activity threshold for
$\mu>0$ shows that the onset of propulsion is not
solely a consequence of periodicity or finite system
size: even on an infinite domain, self-phoretic
feedback must overcome the relaxation of the chemical
field.

The influence of periodicity near onset is controlled
by the ratio of $L$ to the chemical screening length,
$\ell_\mu=\sqrt{D/\mu}$.
When $L\gg\ell_\mu$, the concentration disturbance
decays over distances much shorter than the period,
and the particle responds primarily to its local
chemical environment. The periodic threshold then
approaches the infinite-domain value. Conversely,
when $\ell_\mu$ is comparable to or larger than $L$,
the field extends around the ring, and its periodic
redistribution affects the onset of motion.
The complementary limit $\mu=0$ isolates this
finite-size contribution.

\subsubsection{Zero-evaporation limit}

We next set $\mu=0$ while keeping $L$ finite.
For the positive-velocity branch, the characteristic
roots become
\begin{equation}
    r_+=0,
    \qquad
    r_-=-\frac{v}{D}.
\end{equation}
Using
\begin{equation}
    \lim_{r\to0}
    r\coth\left(\frac{rL}{2}\right)
    =
    \frac{2}{L},
\end{equation}
Eq.~\eqref{eq:single-closed} reduces to
\begin{equation}
    v
    =
    \lambda s
    \left[
        \frac{1}{2D}
        \coth\left(\frac{vL}{2D}\right)
        -
        \frac{1}{Lv}
    \right].
    \label{eq:zero_decay_fixed_point}
\end{equation}
As discussed above, the spatial mean of the concentration
grows when evaporation is absent, but does not contribute
to the gradient driving the particle.
The nonuniform part of the field can nevertheless remain
stationary in the particle frame, allowing constant-velocity
motion. In terms of the dimensionless variables $\widehat v$ and $\lambda^*$, the nonzero
fixed-point condition becomes
\begin{equation}
    y(\widehat v)
    \equiv
    \frac{\coth\widehat v}{\widehat v}
    -
    \frac{1}{\widehat v^2}
    =
    \frac{4}{\lambda^*}.
    \label{eq:zero_decay_dimensionless}
\end{equation}
For $\widehat v>0$, $y(\widehat v)$ decreases monotonically
from $1/3$ at the origin to zero as $\widehat v\to\infty$.
A positive-velocity solution therefore exists if and only if
$4/\lambda^*<1/3$, giving
\begin{equation}
    \lambda^*>\lambda_c^*=12.
    \label{eq:zero_decay_threshold}
\end{equation}
Reflection symmetry supplies the corresponding
negative-velocity solution.
Thus, even without evaporation, diffusion around
the finite periodic domain produces a nonzero activity
threshold. We also note that the scaled speed $|\widehat v|$ compares the
diffusion time across the domain with the particle
traversal time:
\begin{equation}
    2|\widehat v|
    =
    \frac{L^2/D}{L/|v|}.
\end{equation}
Although this ratio has the form of a Péclet number,
the model contains no hydrodynamic advection.
It instead measures how rapidly the chemical source
moves relative to the diffusive redistribution of
its field.

\subsection{Comparison with the one-dimensional channel system}

The infinite-domain limit provides a common reference
for comparing the periodic system with the finite
channel studied in Ref.~\cite{1dchannel}.
The principal distinction between these geometries
is the presence of reflecting boundaries.
In the channel, no-flux boundary conditions modify
the chemical field and break continuous translational
symmetry. The periodic system has no preferred position,
allowing sustained translation around the domain,
whereas motion in the finite channel is constrained
by its boundaries.

\begin{figure}[h]
    \centering
    \includegraphics[width=0.6\textwidth]{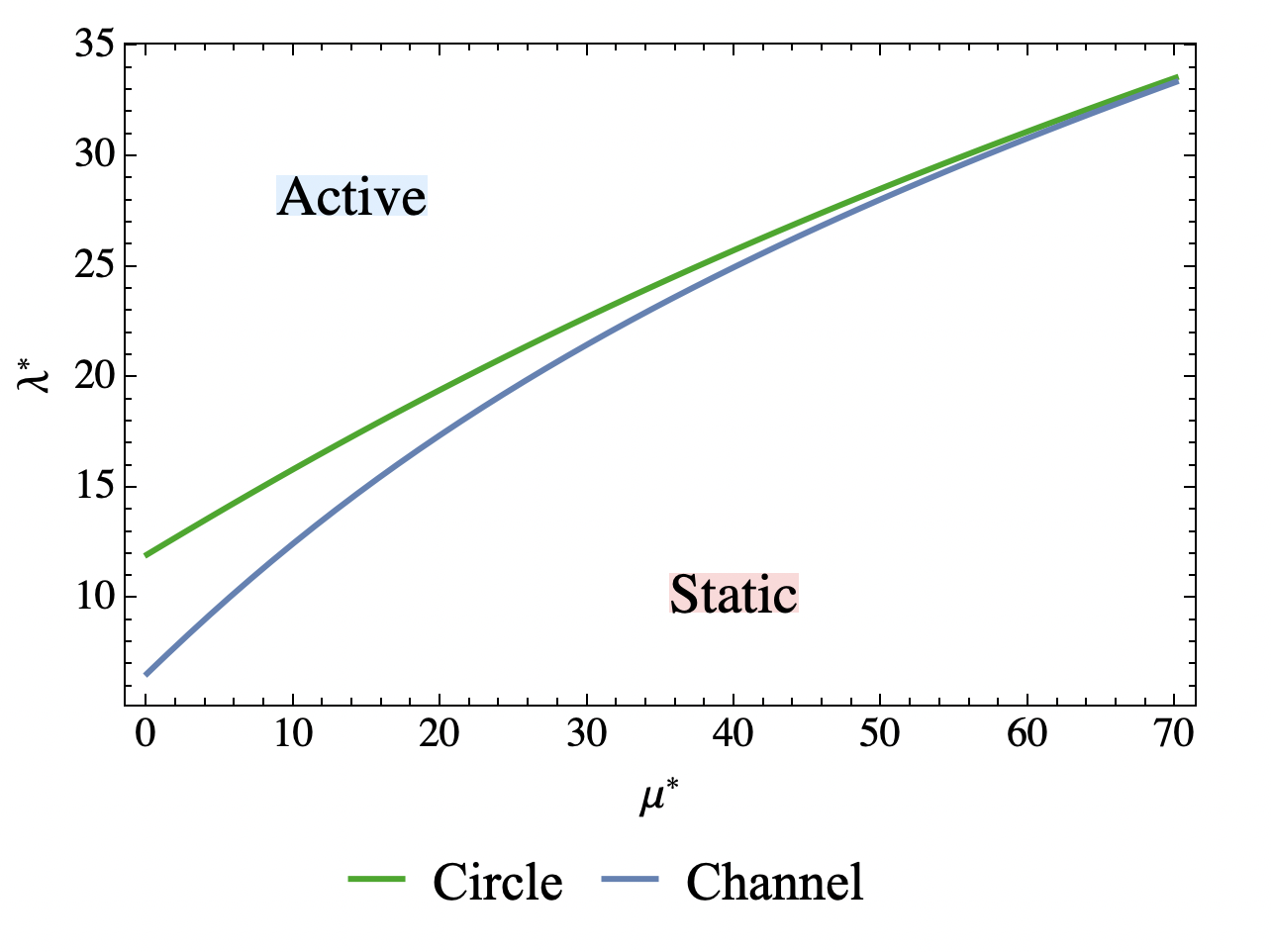}
    \caption{
    Critical activity $\lambda_c^*$ as a function of
    the dimensionless evaporation rate $\mu^*$ for the
    periodic system and the finite channel with
    no-flux boundary conditions studied in
    Ref.~\cite{1dchannel}.
    Both domains have total length $L$.
    The periodic system has a higher threshold at
    small $\mu^*$, while both thresholds approach
    the infinite-domain asymptote
    $\lambda_c^*\sim4\sqrt{\mu^*}$ at large $\mu^*$.
    }
    \label{fig:critical-line-comparison}
\end{figure}

Figure~\ref{fig:critical-line-comparison} compares the
activity thresholds for the two geometries, with both
systems scaled to the same total length $L$.
At small $\mu^*$, the periodic system requires a larger
activity to destabilise the stationary state.
The thresholds approach one another as $\mu^*$
increases and are nearly indistinguishable on the
scale of the figure for $\mu^*\gtrsim60$. 

This dependence on chemical evaporation reflects the spatial
extent of the concentration field. At small $\mu^*$, the chemical field extends across
the domain, making the onset of motion sensitive
to the boundary conditions. The threshold comparison
shows that the stationary state destabilises at a
lower activity in the reflecting channel than in
the periodic system. At large $\mu^*$, the screening length satisfies
$\ell_\mu/L=1/\sqrt{\mu^*}\ll1$.
The chemical field near the stationary particle
is then largely insensitive to distant boundaries
or periodicity, and both thresholds approach the unconfined asymptote
$\lambda_c^*\sim4\sqrt{\mu^*}$, which is obtained by expressing
Eq.~\eqref{eq:infinite_domain_threshold} in dimensionless form. The agreement of the thresholds therefore reflects
a common local mechanism for the onset of motion,
even though the subsequent trajectories remain
constrained by the different geometries.

\subsection{Single-particle numerical results}
\label{subsec:single_particle_numerics}

We now compare the long-time particle velocity obtained
by direct numerical integration, using the method outlined
in Appendix~\ref{sec:numerics}, with the analytical
constant-velocity solution. Unless otherwise stated, the simulations use a Fourier
cutoff $M=256$, timestep $\Delta t^*=10^{-4}$, and final
time $t_{\max}^*=60$.
The steady velocity is estimated by averaging over the
final $25\%$ of the trajectory.
Following the initialisation procedure described in Appendix~\ref{sec:numerics},
a one-step perturbation $U_{\mathrm{seed}}=0.2$ selects
the positive travelling branch.

\begin{figure}[tb]
    \centering
    \includegraphics[width=\textwidth]{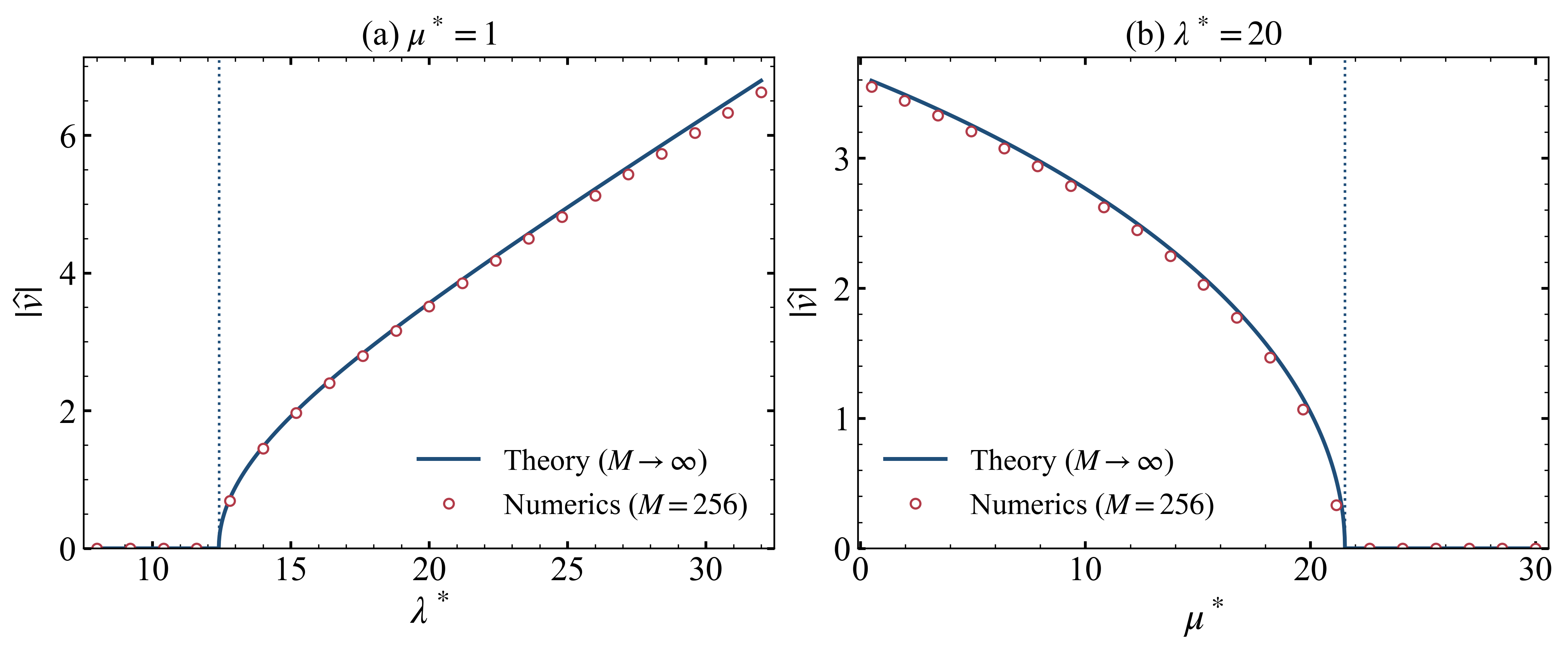}
    \caption{
    Dimensionless steady-state speed
    $|\widehat v|$ as a function of
    (a) mobility $\lambda^*$ at fixed $\mu^*=1$ and
    (b) evaporation rate $\mu^*$ at fixed $\lambda^*=20$.
    Symbols show numerical results with $M=256$,
    $\Delta t^*=10^{-4}$, and $t_{\max}^*=60$,
    averaged over the final $25\%$ of each trajectory.
    Solid curves show the analytical constant-velocity
    solutions in the limit $M\to\infty$.
    Vertical dotted lines mark the corresponding
    analytical transition thresholds.
    }
    \label{fig:single_particle_numerical_results}
\end{figure}

Figure~\ref{fig:single_particle_numerical_results}
compares the numerical and analytical speeds in two
parameter sweeps.
At fixed $\mu^*=1$, increasing the phoretic mobility produces
motion above the analytical threshold
$\lambda_c^*\simeq12.40$, with the speed increasing
continuously along the moving branch.
At fixed $\lambda^*=20$, increasing the chemical evaporation
rate reduces the speed until the moving branch
disappears at $\mu_c^*\simeq21.54$.
Faster chemical evaporation shortens the persistence of the
field through which the particle interacts with its
past trajectory, weakening the feedback that sustains
propulsion.

The numerical results reproduce both trends and agree
closely with the analytical curves, although the
computed speeds are systematically slightly lower.
The threshold values quoted here refer to the
analytical solution; the finite Fourier cutoff,
timestep, and observation time limit the precision
with which onset can be resolved numerically.

\begin{figure}[tb]
    \centering
    \includegraphics[width=\textwidth]{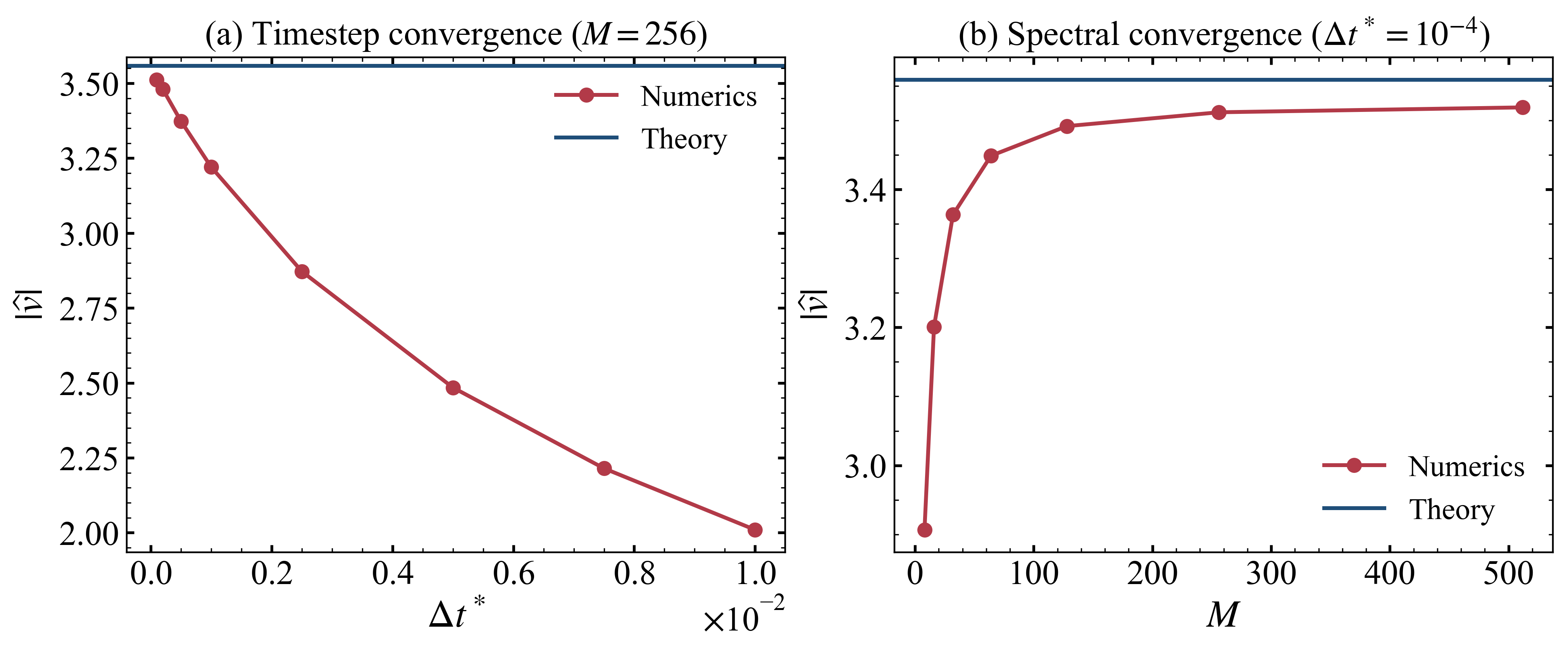}
    \caption{
    Dependence of the numerical steady-state speed
    on (a) the timestep $\Delta t^*$ at fixed $M=256$
    and (b) the Fourier cutoff $M$ at fixed
    $\Delta t^*=10^{-4}$, for $\lambda^*=20$
    and $\mu^*=1$.
    Symbols show numerical results, with connecting
    lines serving as guides to the eye.
    The horizontal line gives the analytical speed
    in the continuum Fourier limit.
    }
    \label{fig:single_particle_convergence}
\end{figure}

To assess discretisation errors, we vary the timestep
and Fourier cutoff separately at $\lambda^*=20$
and $\mu^*=1$.
For the timestep test, $M=256$ is held fixed;
for the Fourier-cutoff test, $\Delta t^*=10^{-4}$
is held fixed.
As shown in Fig.~\ref{fig:single_particle_convergence},
reducing the timestep or increasing the cutoff brings
the numerical speed closer to the analytical value.
Each test retains a residual error associated with
the discretisation held fixed, so neither alone
establishes convergence to the continuum solution.
The results support the use of $M=256$ and
$\Delta t^*=10^{-4}$ for the comparisons above,
while accounting for the small remaining discrepancy
in the measured speeds.

\section{The two-particle system}
\label{sec:two_particle}

With two particles, the chemical field couples individual
propulsion to the evolution of their relative separation.
We first formulate the conditions for states with a fixed
separation and common velocity. We then consider a
non-emitting tracer responding to an active source,
before turning to two mutually coupled emitting particles.

\subsection{General locked-state equations}

Let the particles have source strengths $s_1$ and $s_2$
and mobilities $\lambda_1$ and $\lambda_2$.
For a uniformly translating locked state,
$X_i(t)=Y_i+vt$, translational invariance allows us to
choose $Y_1=0$ and write
\begin{equation}
    d_{12}=Y_2-Y_1,
    \qquad 0<d_{12}<L.
\end{equation}
The directed separation $d_{12}$ and common velocity
$v$ satisfy
\begin{align}
    v
    &=
    \lambda_1
    \left[
        s_1K_\mu^{(1)}(0,0;v)
        +
        s_2K_\mu^{(1)}(0,d_{12};v)
    \right],
    \label{eq:two-locked-1}
    \\
    v
    &=
    \lambda_2
    \left[
        s_1K_\mu^{(1)}(d_{12},0;v)
        +
        s_2K_\mu^{(1)}(0,0;v)
    \right],
    \label{eq:two-locked-2}
\end{align}
where translational invariance gives
$K_\mu^{(1)}(d_{12},d_{12};v)=K_\mu^{(1)}(0,0;v)$.
Each equation combines the response to the particle's
own chemical field with its response to the field
of the other particle. Together, they determine
$v$ and $d_{12}$ whenever a locked state exists.
The case $v=0$ describes a stationary configuration.
The linear response about such a locked state is governed
by the $2\times2$ inverse response matrix
\begin{equation}
    \widetilde{\boldsymbol S}(\omega)
    =
    \begin{pmatrix}
        \widetilde S_{11}(\omega)
        &
        \widetilde S_{12}(\omega)
        \\
        \widetilde S_{21}(\omega)
        &
        \widetilde S_{22}(\omega)
    \end{pmatrix}.
\end{equation}
Specialising Eq.~\eqref{eq:inverse_response_matrix}
to two particles gives
\begin{align}
\widetilde S_{11}(\omega)
&=
\ii\omega
-\lambda_1\left[
    s_1K_\mu^{(2)}(0,0;v)
    +s_2K_{\mu}^{(2)}(0,d_{12};v)
\right]
+\lambda_1s_1K_{\mu+\ii\omega}^{(2)}(0,0;v),
\\
\widetilde S_{12}(\omega)
&=
\lambda_1s_2K_{\mu+\ii\omega}^{(2)}(0,d_{12};v),
\\
\widetilde S_{21}(\omega)
&=
\lambda_2s_1K_{\mu+\ii\omega}^{(2)}(d_{12},0;v),
\\
\widetilde S_{22}(\omega)
&=
\ii\omega
-\lambda_2\left[
    s_1K_{\mu}^{(2)}(d_{12},0;v)
    +s_2K_{\mu}^{(2)}(0,0;v)
\right]
+\lambda_2s_2K_{\mu+\ii\omega}^{(2)}(0,0;v).
\end{align}
The stability of a locked state is then determined by the
zeros of $\det\widetilde{\boldsymbol S}(\omega)$.
Candidate thresholds for stationary and oscillatory
instabilities satisfy Eqs.~\eqref{eq:pitch} and
\eqref{eq:hopf-condition}, respectively.

\subsection{Active source coupled to an inactive tracer}
\label{subsec:active_tracer}

We first consider a single chemical source and a
non-emitting tracer, setting
\begin{equation}
    s_1=s,
    \qquad
    s_2=0.
\end{equation}
Although the tracer produces no chemical, it responds
to concentration gradients when $\lambda_2\neq0$.
The coupling is therefore unidirectional: the source
drives the tracer through its chemical field, while
the tracer has no influence on the source. The source 
consequently obeys the single-particle
dynamics studied in Sec.~\ref{sec:single_particle}.
We consider it on a constant-velocity branch with
velocity $v$, generating a travelling concentration
profile $c(x,t)$.
In the laboratory frame, the tracer obeys
\begin{equation}
    \dot X_2(t)
    =
    -\lambda_2\partial_x c\bigl(X_2(t),t\bigr).
    \label{eq:active_tracer_equation}
\end{equation}
The remaining dynamics are determined by the tracer's
motion relative to the travelling source. Depending on
the strength and sign of its phoretic response, $\lambda_2$, the
tracer may either approach a fixed separation from the source
or continually drift relative to the source and repeatedly circulate around the ring. We refer to these distinct dynamical states as locking and slipping, respectively.

\subsubsection{Relative dynamics and existence of locked states}

To describe the tracer motion in the source frame,
we introduce the directed separation
\begin{equation}
    \Delta(t)=X_2(t)-X_1(t),
\end{equation}
understood modulo $L$.
Writing $c(x,t)=c_v(x-X_1(t))$ and using
$\dot X_1=v$, Eq.~\eqref{eq:active_tracer_equation}
becomes
\begin{equation}
    \dot\Delta
    =
    -\lambda_2c_v'(\Delta)-v
    \equiv F_v(\Delta).
    \label{eq:active_tracer_separation_dynamics}
\end{equation}
The tracer dynamics thus reduce to an autonomous
one-dimensional equation: its phoretic velocity
relative to the laboratory frame competes with
the translation of the source. A locked state then corresponds to a fixed point,
\begin{equation}
    \Delta(t)=d_{12},
    \qquad
    F_v(d_{12})=0,
    \qquad
    0<d_{12}<L.
\end{equation}
The locking condition is therefore
\begin{equation}
    v=-\lambda_2c_v'(d_{12}),
    \label{eq:active_tracer_locking_condition}
\end{equation}
which requires the tracer's phoretic velocity to
match the source velocity.
This is Eq.~\eqref{eq:two-locked-2} specialised to
$s_1=s$ and $s_2=0$; Eq.~\eqref{eq:two-locked-1}
independently determines $v$ through the
single-particle self-consistency condition.
For $\lambda_2\neq0$, we may equivalently write
\begin{equation}
    c_v'(d_{12})=-\frac{v}{\lambda_2}.
    \label{eq:active_tracer_graphical_condition}
\end{equation}
A noncoincident locked state therefore exists
wherever the horizontal level $-v/\lambda_2$
intersects the concentration gradient $c_v'(x)$
for $x\in(0,L)$. The concentration profile is continuous at the source
but has a cusp there, so its spatial derivative has a
jump discontinuity. Integrating the steady advection--diffusion
equation across the source gives
\begin{equation}
    c_v'(0^+)-c_v'(L^-)
    =
    -\frac{s}{D}.
    \label{eq:active_tracer_gradient_jump}
\end{equation}
The source itself responds to the symmetric average of
the one-sided gradients,
\begin{equation}
    c_{v,\mathrm{reg}}'(0)
    =
    \frac{1}{2}
    \left[
        c_v'(0^+)+c_v'(L^-)
    \right].
\end{equation}
Since the source travels with velocity $v$, its equation
of motion gives
\begin{equation}
    v=-\lambda_1c_{v,\mathrm{reg}}'(0),
    \label{eq:cvreg}
\end{equation}
and hence
\begin{equation}
    c_v'(0^+)+c_v'(L^-)
    =
    -\frac{2v}{\lambda_1}.
    \label{eq:active_tracer_gradient_sum}
\end{equation}
Solving Eqs.~\eqref{eq:active_tracer_gradient_jump}
and \eqref{eq:active_tracer_gradient_sum} yields
\begin{align}
    c_v'(0^+)
    &=
    -\frac{v}{\lambda_1}
    -\frac{s}{2D},
    \label{eq:active_tracer_cusp_plus}
    \\
    c_v'(L^-)
    =
    c_v'(0^-)
    &=
    -\frac{v}{\lambda_1}
    +\frac{s}{2D}.
    \label{eq:active_tracer_cusp_minus}
\end{align}

For $s>0$, the gradient is continuous on the open
interval $(0,L)$ and spans the values between these
one-sided limits. The intermediate value theorem
therefore guarantees an intersection with the level
$-v/\lambda_2$ whenever
\begin{equation}
    c_v'(0^+)
    <
    -\frac{v}{\lambda_2}
    <
    c_v'(L^-).
    \label{eq:active_tracer_ivt_condition}
\end{equation}
Substituting Eqs.~\eqref{eq:active_tracer_cusp_plus}
and \eqref{eq:active_tracer_cusp_minus} gives
\begin{equation}
    \left|
        v\left(
            \frac{1}{\lambda_1}
            -
            \frac{1}{\lambda_2}
        \right)
    \right|
    <
    \frac{|s|}{2D}.
    \label{eq:active_tracer_existence_condition}
\end{equation}
As shown in
Appendix~\ref{app:active_tracer_monotonicity},
$c_v'(x)$ is monotonic on $(0,L)$. Consequently,
Eq.~\eqref{eq:active_tracer_existence_condition} is
both necessary and sufficient for the existence of a
noncoincident locked state, and the corresponding
separation is unique. If this condition is violated,
$F_v(\Delta)$ has no zero and the tracer remains in
the slipping state. The existence condition in
Eq.~\eqref{eq:active_tracer_existence_condition}
generally defines a region in the
$(\lambda_1,\lambda_2)$ parameter plane, rather than
a single lower threshold in $\lambda_2$. Its form
depends on the signs of the source velocity
and the two mobilities. A single upper or lower
threshold emerges only after these signs and the
relevant boundary branch have been specified.

\subsubsection{Boundaries of the locking region}

At the boundaries of the locking region, the fixed
point approaches the source from one side:
$d_{12}\to0^+$ or $d_{12}\to L^-\equiv0^-$.
The two boundary conditions follow by saturating
Eq.~\eqref{eq:active_tracer_existence_condition},
\begin{equation}
    v\left(
        \frac{1}{\lambda_1}
        -
        \frac{1}{\lambda_2}
    \right)
    =
    \pm\frac{s}{2D}.
    \label{eq:active_tracer_boundary_implicit}
\end{equation}
For $v>0$ and $d_{12}=X_2-X_1$, the negative-sign
branch corresponds to $d_{12}\to0^+$, with the tracer
immediately ahead of the source. The positive-sign
branch corresponds to $d_{12}\to L^-$, with the tracer
immediately behind it. To obtain a parametric representation of these
boundaries, we use the regularised self-gradient
\begin{equation}
\begin{aligned}
    A(v)
    &\equiv c_{v,\mathrm{reg}}'(0)
    \\
    &=
    \frac{s}{2D(r_+-r_-)}
    \left[
        r_+\coth\left(\frac{r_+L}{2}\right)
        -
        r_-\coth\left(\frac{r_-L}{2}\right)
    \right],
\end{aligned}
\end{equation}
which is obtained from the single-particle travelling profile derived in Eq.~\eqref{eq:single-closed}. Equations~\eqref{eq:cvreg} and
\eqref{eq:active_tracer_boundary_implicit} then give
\begin{align}
    \lambda_1(v)
    &=
    -\frac{v}{A(v)},
    \label{eq:active_tracer_boundary_parametric_lambda1}
    \\
    \lambda_2^\pm(v)
    &=
    \left[
        \frac{1}{\lambda_1(v)}
        \mp\frac{s}{2Dv}
    \right]^{-1}.
    \label{eq:active_tracer_boundary_parametric_lambda2}
\end{align}
Here, $\lambda_2^-(v)$ denotes the boundary approached
with the tracer ahead of the source, while
$\lambda_2^+(v)$ denotes the boundary approached from
behind.

\subsubsection{Stability of the locked solution}

Here we analyse the stability of the locked solution. We perturb the
separation about its locked value:
\begin{equation}
    \Delta(t)=d_{12}+\delta\Delta(t),
    \qquad
    |\delta\Delta(t)|\ll1.
\end{equation}
Linearising
Eq.~\eqref{eq:active_tracer_separation_dynamics}
gives
\begin{equation}
\begin{aligned}
    \delta\dot\Delta(t)
    &=
    F_v'(d_{12})\,\delta\Delta(t)
    \\
    &=
    -\lambda_2c_v''(d_{12})\,
    \delta\Delta(t).
\end{aligned}
\label{eq:active_tracer_linearised_relative_dynamics}
\end{equation}
The perturbation therefore evolves as
\begin{equation}
    \delta\Delta(t)
    =
    \delta\Delta(0)
    \exp\left[
        -\lambda_2c_v''(d_{12})t
    \right].
\end{equation}
The locked separation is linearly stable when
\begin{equation}
    \lambda_2c_v''(d_{12})>0,
    \label{eq:active_tracer_stability_condition}
\end{equation}
and unstable when the inequality is reversed.
As shown in
Appendix~\ref{app:active_tracer_monotonicity},
$c_v''(x)$ has the same sign as $s$ throughout
$0<x<L$. The stability condition consequently
reduces to
\begin{equation}
    \boxed{\lambda_2s>0.}
    \label{eq:active_tracer_stability_sign}
\end{equation}

We henceforth restrict attention to $s>0$ and
$\lambda_2>0$, for which the tracer moves down the
concentration gradient and is repelled by the source.
For a source travelling with $v>0$, the locking
condition
\begin{equation}
    c_v'(d_{12})=-\frac{v}{\lambda_2}<0
\end{equation}
places the stable tracer ahead of the source.
The boundary of the locking region is reached as
$d_{12}\to0^+$ and is therefore
\begin{equation}
    \lambda_{2,\mathrm{lock}}(v)
    =
    \left[
        \frac{1}{\lambda_1(v)}
        +
        \frac{s}{2Dv}
    \right]^{-1}.
    \label{eq:active_tracer_ahead_boundary}
\end{equation}

\subsubsection{Large-source-mobility limit}

The form of the stable locking boundary becomes
particularly simple when the source mobility is large,
$\lambda_1\gg\lambda_{1c}$. In this regime, the source
velocity is also large compared with the chemical
diffusion--evaporation scale, 
\begin{equation}
    v\gg\sqrt{\mu D}.
\end{equation}
The characteristic roots then satisfy
\begin{equation}
    r_+\simeq\frac{\mu}{v},
    \qquad
    r_-\simeq-\frac{v}{D}.
\end{equation}
Thus, $r_+\to0$ while $r_-\to-\infty$ as
$v\to\infty$. The concentration gradient in the source frame is
\begin{equation}
    c_v'(x)
    =
    \frac{s}{D(r_+-r_-)}
    \left[
        \frac{r_-e^{r_-x}}{1-e^{r_-L}}
        +
        \frac{r_+e^{r_+(x-L)}}{1-e^{-r_+L}}
    \right].
\end{equation}
Substituting the large-velocity forms of $r_\pm$ gives
the one-sided limits
\begin{equation}
    c_v'(0^+)\simeq-\frac{s}{D},
    \qquad
    c_v'(L^-)\simeq0.
\end{equation}
The chemical field therefore develops a strong gradient
immediately ahead of the moving source, while the gradient
immediately behind it becomes negligible. The regularised
self-gradient approaches
\begin{equation}
    c_{v,\mathrm{reg}}'(0)
    \simeq-\frac{s}{2D}.
\end{equation}
The source self-consistency condition then gives
\begin{equation}
    v\simeq\frac{\lambda_1s}{2D}.
    \label{eq:active_tracer_large_lambda1_velocity}
\end{equation}
In the same limit, the gradient spans approximately
\begin{equation}
    -\frac{s}{D}
    \lesssim c_v'(x)\lesssim0.
\end{equation}
For $s>0$, $v>0$, and $\lambda_2>0$, the locking
condition can therefore be satisfied only if
\begin{equation}
    \frac{v}{\lambda_2}
    \lesssim\frac{s}{D},
\end{equation}
or equivalently,
\begin{equation}
    \lambda_2\gtrsim\frac{Dv}{s}.
\end{equation}
Using Eq.~\eqref{eq:active_tracer_large_lambda1_velocity}
yields the asymptotic locking boundary
\begin{equation}
    \boxed{
        \lambda_2\gtrsim\frac{\lambda_1}{2}.
    }
    \label{eq:active_tracer_large_lambda1_boundary}
\end{equation}
Thus, a rapidly moving source can entrain the tracer only
if the tracer mobility increases proportionally with the
source mobility. The factor \(1/2\) reflects the limiting
self-gradient experienced by the source: the source
responds to the symmetric average across the cusp,
whereas a tracer immediately ahead experiences the full
one-sided gradient.

\subsubsection{Source--tracer phase diagram}

\begin{figure}[tb]
    \centering
    \includegraphics[width=0.8\textwidth]{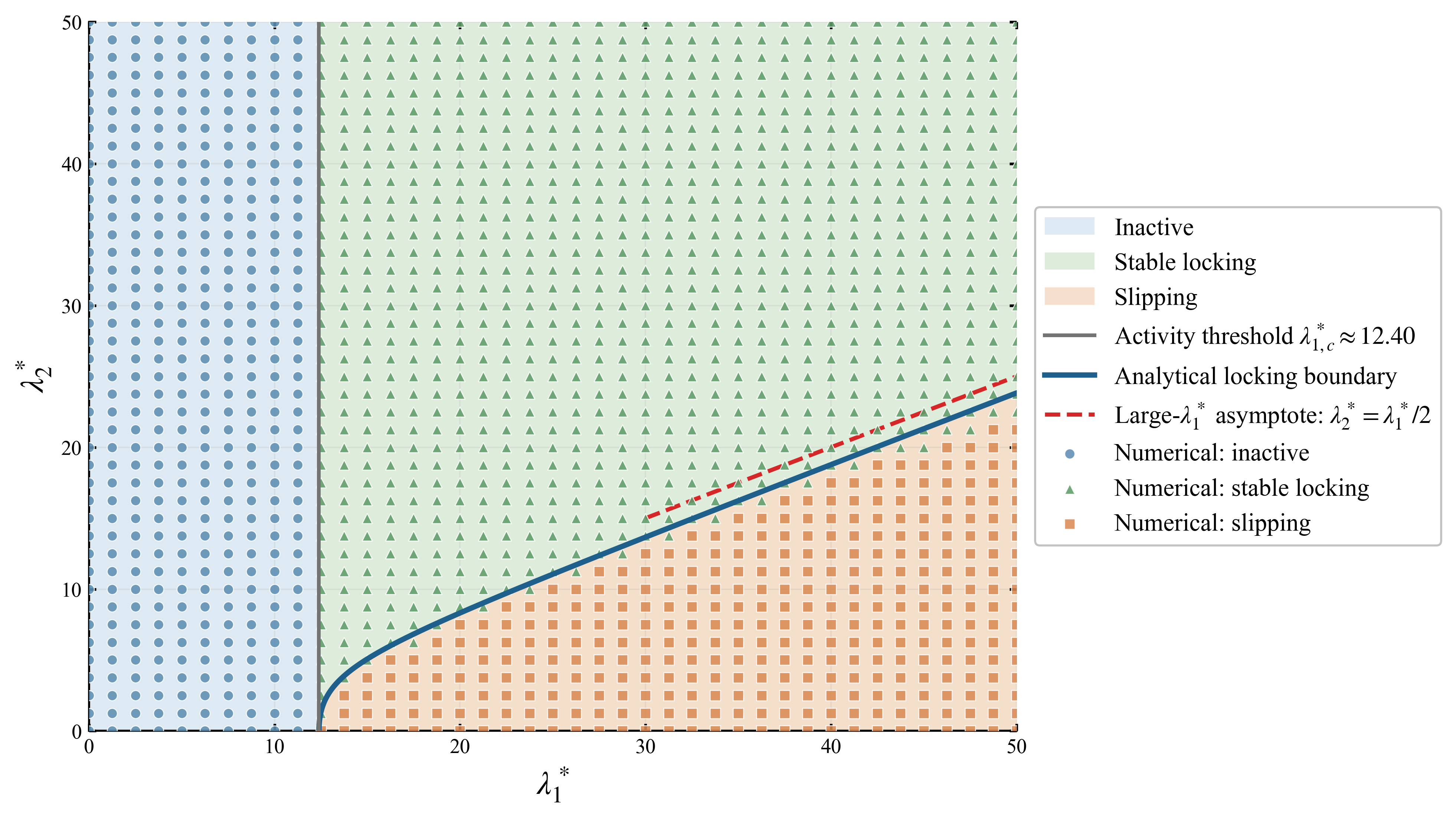}
    \caption{
    Source--tracer dynamics in the stable sector
    $\lambda_2^*>0$ for $\mu^*=1$.
    The shaded regions denote a stationary source
    (blue), stable locking (green), and slipping
    (orange).
    The vertical grey line marks the single-particle
    activity threshold
    $\lambda_{1,c}^*\simeq12.40$, and the solid blue
    curve is the analytical slipping--locking boundary.
    The red dashed line shows the large-mobility
    asymptote $\lambda_2^*=\lambda_1^*/2$.
    Symbols show the numerical classification on a
    $41\times41$ parameter grid: stationary source
    (circles), slipping (squares), and stable locking
    (triangles).
    }
    \label{fig:active_tracer_phase_diagram}
\end{figure}

Figure~\ref{fig:active_tracer_phase_diagram}
summarises the source--tracer dynamics in the
$(\lambda_1^*,\lambda_2^*)$ plane for $\mu^*=1$.
Below the single-particle activity threshold,
$\lambda_1^*<\lambda_{1,c}^*\simeq12.40$, the source
remains stationary. Once the source begins to move,
the tracer either slips relative to it or becomes
entrained at a fixed separation. For $\lambda_2^*>0$, the locked separation is stable
and the tracer travels ahead of the source. Below the
analytical locking boundary, the tracer mobility is
insufficient for it to match the source velocity, and
the tracer repeatedly slips around the ring. Above the
boundary, the two particles travel with the same
velocity while maintaining a fixed separation.
At large source mobility, the boundary approaches
$\lambda_2^*=\lambda_1^*/2$, in agreement with
Eq.~\eqref{eq:active_tracer_large_lambda1_boundary}.

To test these predictions, we performed direct numerical
simulations using the method outlined in Appendix~\ref{sec:numerics} on a $41\times41$ uniform grid spanning
$0\leq\lambda_1^*,\lambda_2^*\leq50$.
The simulations used $M=256$ and a baseline timestep
$\Delta t^*=10^{-4}$, reduced to $10^{-5}$ close to
the locking boundary at large $\lambda_1^*$.
The final time was increased from $t_{\max}^*=60$ to
$100$--$300$ near the activity and locking boundaries
to allow for slow relaxation. After identifying simulations with a stationary source,
an active trajectory was classified as locked when
\begin{equation}
    \left|
        \left\langle U_2-U_1\right\rangle
    \right|
    <10^{-3}
\end{equation}
and the unwrapped relative displacement varied by less
than $0.1L$ over the measurement interval. Trajectories
that continued to accumulate relative displacement were
classified as slipping. The numerical classifications reproduce all three
predicted regimes and closely follow both the
single-particle activity threshold and the analytical
slipping--locking boundary. In particular, the results
show that a non-emitting tracer can be entrained by a
travelling chemical source, provided that its mobility
is sufficiently large relative to that of the source.

\subsection{Two interacting active sources}
\label{subsec:twosources}

We now consider two emitting particles,
\begin{equation}
    s_1,s_2>0,
    \qquad
    \lambda_1,\lambda_2>0.
\end{equation}
Unlike the source--tracer system, the coupling is now
mutual: each particle modifies the chemical field that
drives both its own motion and that of the other particle.
The resulting dynamics therefore cannot be reduced to
the motion of a passive degree of freedom in a prescribed
travelling field. For two separated particles,
Eq.~\eqref{eq:stationary_spacing_condition} admits the
stationary configuration
\begin{equation}
    d_{12}=d_{21}=\frac{L}{2}.
    \label{eq:two_source_antipodal_state}
\end{equation}
The particles are therefore antipodal in the stationary
state, irrespective of their relative source strengths
and mobilities. These parameters determine the stability
of the stationary configuration, but not its separation.

\subsubsection{Competing instabilities of the antipodal state}

The linear modes of the antipodal state are determined
by the zeros of
$\det\widetilde{\boldsymbol S}(\omega)$.
Two distinct mechanisms can then destabilise this inactive rest phase.
Firstly, a stationary instability may occur if an additional
zero reaches $\omega=0$. In the present system, this
marks the onset of uniformly translating locked states,
in which the particles move with a common velocity while
maintaining a fixed separation. Reflection symmetry
relates solutions travelling in opposite directions,
giving the two branches of a pitchfork bifurcation.
Alternatively, a pair of zeros may reach the real \(\omega\)-axis at opposite nonzero frequencies,
\begin{equation}
    \omega=\pm\omega_{\mathrm H},
    \qquad
    \omega_{\mathrm H}>0.
\end{equation}
This signals a Hopf instability of the antipodal state,
in which perturbations to the particle configuration
become oscillatory. The following analysis determines which instability
is encountered first as the coupling strength is
increased. It identifies the nature and frequency
of the critical linear mode, while the nonlinear
dynamics beyond onset are examined numerically below.

\subsubsection{Linear stability of the stationary antipodal state}

We now specialise the response matrix to the stationary
antipodal configuration,
\begin{equation}
    v=0,
    \qquad
    d_{12}=\frac{L}{2}.
\end{equation}
In this subsection, we work in dimensionless units with
$L=D=1$ and omit the stars from the dimensionless
parameters for notational simplicity. The determinant
of the inverse response matrix can then
be written as
\begin{align}
    \det\widetilde{\boldsymbol S}(\omega)
    ={}&
    -\omega^2
    +(\lambda_1s_1+\lambda_2s_2)\mathcal A(\omega)
    +(\lambda_1s_2+\lambda_2s_1)\mathcal B(\omega)
    \nonumber\\
    &+
    \lambda_1\lambda_2s_1s_2\mathcal C(\omega)
    +
    \lambda_1\lambda_2(s_1^2+s_2^2)\mathcal D(\omega),
    \label{eq:two_source_stationary_determinant}
\end{align}
where
\begin{align}
    \mathcal A(\omega)
    ={}&
    \frac{\ii\omega}{2}
    \left[
        \sqrt{\mu}\,
        \coth\left(\frac{\sqrt{\mu}}{2}\right)
        -
        \sqrt{\mu+\ii\omega}\,
        \coth\left(
            \frac{\sqrt{\mu+\ii\omega}}{2}
        \right)
    \right],
    \\
    \mathcal B(\omega)
    ={}&
    \frac{\ii\omega\sqrt{\mu}}{2}
    \csch\left(\frac{\sqrt{\mu}}{2}\right),
    \\
    \mathcal C(\omega)
    ={}&
    \frac{\ii\omega}{4}
    +
    \frac{\sqrt{\mu}}{2}
    \coth\left(\frac{\sqrt{\mu}}{2}\right)
    \left[
        \sqrt{\mu}\,
        \coth\left(\frac{\sqrt{\mu}}{2}\right)
        -
        \sqrt{\mu+\ii\omega}\,
        \coth\left(
            \frac{\sqrt{\mu+\ii\omega}}{2}
        \right)
    \right],
    \\
    \mathcal D(\omega)
    ={}&
    \frac{\sqrt{\mu}}{4}
    \csch\left(\frac{\sqrt{\mu}}{2}\right)
    \left[
        \sqrt{\mu}\,
        \coth\left(\frac{\sqrt{\mu}}{2}\right)
        -
        \sqrt{\mu+\ii\omega}\,
        \coth\left(
            \frac{\sqrt{\mu+\ii\omega}}{2}
        \right)
    \right].
\end{align}
The derivation is given in
Appendix~\ref{app:two_source_response_determinant}.
All four functions vanish at $\omega=0$, making the
neutral translational zero of the determinant explicit.
A stationary instability occurs when an additional zero
reaches the origin after this neutral factor is removed.

To separate the overall interaction strength from the
asymmetry between the particles, we introduce
\begin{equation}
    s_1=s\cos\theta,
    \qquad
    s_2=s\sin\theta,
    \qquad
    \lambda_1=\lambda\cos\phi,
    \qquad
    \lambda_2=\lambda\sin\phi,
    \label{eq:angular_parametrisation}
\end{equation}
with
\begin{equation}
    0<\theta,\phi<\frac{\pi}{2},
\end{equation}
and define
\begin{equation}
    g=\lambda s.
\end{equation}
Here, $s$ and $\lambda$ set the overall scales of the
source strengths and mobilities, while $\theta$ and
$\phi$ describe their relative values.
Equal source strengths correspond to $\theta=\pi/4$,
and equal mobilities to $\phi=\pi/4$; identical
particles lie at their intersection. The endpoints
$\theta,\phi\in\{0,\pi/2\}$ correspond to limiting
cases in which one source strength or mobility vanishes.
Substitution into $\det\widetilde{\boldsymbol S}(\omega)   = 0$
reduces the characteristic equation to a quadratic
in the overall coupling strength, $g$:
\begin{equation}
    \mathcal Q(\omega;\theta,\phi)g^2
    +
    \mathcal L(\omega;\theta,\phi)g
    -
    \omega^2
    =0,
    \label{eq:complex_quadratic}
\end{equation}
where
\begin{align}
    \mathcal L(\omega;\theta,\phi)
    &=
    \mathcal A(\omega)\cos(\theta-\phi)
    +
    \mathcal B(\omega)\sin(\theta+\phi),
    \label{eq:def_L}
    \\
    \mathcal Q(\omega;\theta,\phi)
    &=
    \frac{\sin(2\phi)}{4}
    \left[
        \mathcal C(\omega)\sin(2\theta)
        +
        2\mathcal D(\omega)
    \right].
    \label{eq:def_Q}
\end{align}

\paragraph{Hopf threshold.}

At a Hopf instability, a pair of zeros reaches the real
$\omega$-axis at a nonzero frequency. We therefore seek
solutions of Eq.~\eqref{eq:complex_quadratic} satisfying
\begin{equation}
    g>0,
    \qquad
    \omega\in\mathbb R\setminus\{0\}.
\end{equation}
Writing
\begin{equation}
    \mathcal L
    =
    \mathcal L_{\mathrm R}
    +
    \ii\mathcal L_{\mathrm I},
    \qquad
    \mathcal Q
    =
    \mathcal Q_{\mathrm R}
    +
    \ii\mathcal Q_{\mathrm I},
\end{equation}
the imaginary part of
Eq.~\eqref{eq:complex_quadratic} gives
\begin{equation}
    g(\omega)
    =
    -\frac{\mathcal L_{\mathrm I}(\omega)}
    {\mathcal Q_{\mathrm I}(\omega)},
    \label{eq:hopf_g_of_omega}
\end{equation}
provided that
$\mathcal Q_{\mathrm I}(\omega)\neq0$.
Substituting this expression into the real part yields
the scalar frequency equation
\begin{equation}
    \mathcal Q_{\mathrm R}(\omega)
    \frac{\mathcal L_{\mathrm I}(\omega)^2}
    {\mathcal Q_{\mathrm I}(\omega)^2}
    -
    \mathcal L_{\mathrm R}(\omega)
    \frac{\mathcal L_{\mathrm I}(\omega)}
    {\mathcal Q_{\mathrm I}(\omega)}
    -
    \omega^2
    =0.
    \label{eq:hopf_frequency_condition}
\end{equation}
For specified $(\theta,\phi)$, this equation is solved
numerically for positive frequencies. Each admissible
solution $\omega_{\mathrm H}>0$ gives a candidate
threshold
\begin{equation}
    g_{\mathrm H}
    =
    -\frac{\mathcal L_{\mathrm I}(\omega_{\mathrm H})}
    {\mathcal Q_{\mathrm I}(\omega_{\mathrm H})}.
    \label{eq:hopf_threshold}
\end{equation}
If several admissible solutions exist, the relevant
Hopf threshold is the smallest positive value of
$g_{\mathrm H}$ encountered as the coupling is increased
from the stable regime.

\paragraph{Pitchfork threshold.}

For a stationary instability, we instead evaluate the
criticality condition as $\omega\to0$.
After division by the neutral factor $\ii\omega$, the
term $-\omega^2$ vanishes in this limit, giving
\begin{equation}
    g_{\mathrm P}
    =
    -\frac{4}{\sin(2\phi)}
    \lim_{\omega\to0}
    \frac{
        \mathcal A(\omega)\cos(\theta-\phi)
        +
        \mathcal B(\omega)\sin(\theta+\phi)
    }{
        \mathcal C(\omega)\sin(2\theta)
        +
        2\mathcal D(\omega)
    }.
    \label{eq:pitchfork_zero_frequency_limit}
\end{equation}
Both the numerator and denominator vanish at the origin,
so we retain their leading small-frequency terms:
\begin{align}
    \mathcal A(\omega)
    &=
    O(\omega^2),
    \\
    \mathcal B(\omega)
    &=
    \ii\omega b_1,
    \\
    \mathcal C(\omega)
    &=
    \ii\omega c_1+O(\omega^2),
    \\
    \mathcal D(\omega)
    &=
    \ii\omega d_1+O(\omega^2),
\end{align}
where
\begin{align}
    b_1
    &=
    \frac{\sqrt{\mu}}{2}
    \csch\left(\frac{\sqrt{\mu}}{2}\right),
    \\
    c_1
    &=
    \frac{1}{8}
    \csch^2\left(\frac{\sqrt{\mu}}{2}\right)
    \left[
        \sqrt{\mu}\,
        \coth\left(\frac{\sqrt{\mu}}{2}\right)
        -2
    \right],
    \\
    d_1
    &=
    \frac{1}{16}
    \csch\left(\frac{\sqrt{\mu}}{2}\right)
    \left[
        \sqrt{\mu}\,
        \csch^2\left(\frac{\sqrt{\mu}}{2}\right)
        -
        2\coth\left(\frac{\sqrt{\mu}}{2}\right)
    \right].
\end{align}
The function $\mathcal A$ does not contribute at leading
order, while the common factor $\ii\omega$ cancels from
the remaining terms. Hence,
\begin{equation}
    g_{\mathrm P}
    =
    -\frac{4}{\sin(2\phi)}
    \frac{
        b_1\sin(\theta+\phi)
    }{
        c_1\sin(2\theta)+2d_1
    }.
\end{equation}
Substitution of the coefficients gives
\begin{equation}
    g_{\mathrm P}(\theta,\phi)
    =
    \frac{
        8\sqrt{\mu}\,
        \sinh^2\left(\frac{\sqrt{\mu}}{2}\right)
        \sin(\theta+\phi)
    }{
        \sin(2\phi)
        \left[
            \sinh\left(\frac{\sqrt{\mu}}{2}\right)
            \cosh\left(\frac{\sqrt{\mu}}{2}\right)
            -
            \frac{\sqrt{\mu}}{2}
            +
            \sin(2\theta)
            \left(
                \sinh\left(\frac{\sqrt{\mu}}{2}\right)
                -
                \frac{\sqrt{\mu}}{2}
                \cosh\left(\frac{\sqrt{\mu}}{2}\right)
            \right)
        \right]
    }.
    \label{eq:pitchfork_threshold_angular}
\end{equation}
The first instability is then determined by comparing the
admissible positive thresholds as the coupling \(g\) is
increased from the stable regime. The critical coupling is
\begin{equation}
    g_c^{m}(\theta,\phi)
    =
    \min\left\{
        g_{\mathrm P}(\theta,\phi),
        g_{\mathrm H}(\theta,\phi)
    \right\},
    \label{eq:two_source_first_threshold}
\end{equation}
where the minimum is taken over the available finite positive
thresholds. Thus, if
$g_{\mathrm P}<g_{\mathrm H}$, the antipodal state first
loses stability through the pitchfork mode associated
with uniformly translating locked states. If
$g_{\mathrm H}<g_{\mathrm P}$, it first loses stability
through a finite-frequency Hopf mode. The locus
$g_{\mathrm P}=g_{\mathrm H}$ separates these two
linear instability mechanisms in the
$(\theta,\phi)$ parameter plane.

\begin{figure}[h]
    \centering

    \begin{subfigure}[t]{0.47\textwidth}
        \centering
        \includegraphics[width=\linewidth]{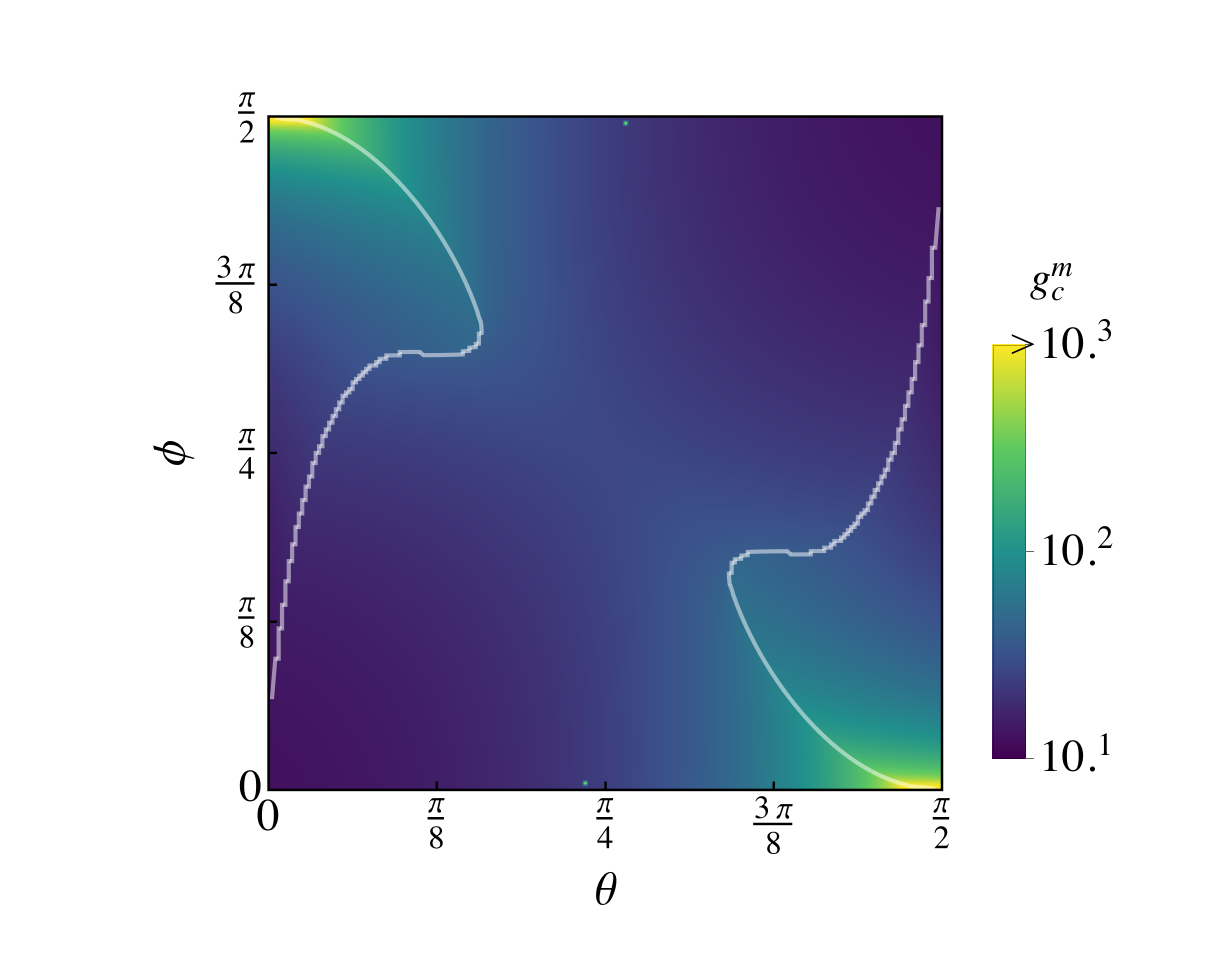}
        \caption{
        Critical coupling \(g_c^{m}(\theta,\phi)\), shown on a
        logarithmic colour scale. The white contours indicate the locus
        \(g_{\mathrm P}=g_{\mathrm H}\), with some visible stair-stepping due to the finite angular grid. 
        }
        \label{fig:two_particle_threshold}
    \end{subfigure}
    \hfill
    \begin{subfigure}[t]{0.5\textwidth}
        \centering
        \includegraphics[width=\linewidth]{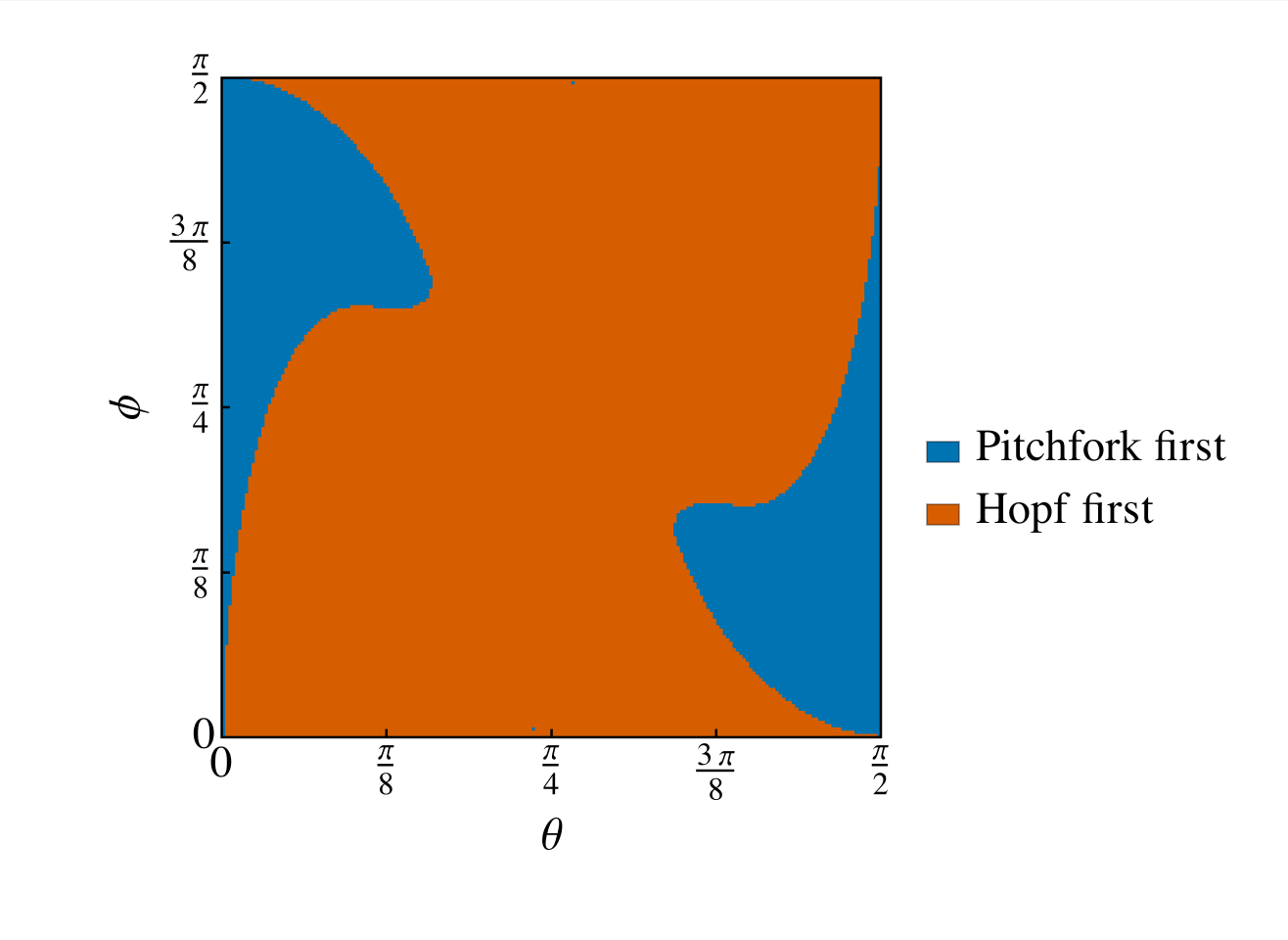}
        \caption{
        Nature of the first instability. Blue and orange denote
        pitchfork-first and Hopf-first regions, respectively.
        }
        \label{fig:two_particle_transition}
    \end{subfigure}

    \caption{
    Linear stability of the stationary antipodal state over the
    \((\theta,\phi)\) parameter plane. The angles \(\theta\) and
    \(\phi\) determine the relative source strengths and phoretic
    response coefficients, respectively. Boundary cases in which
    one source strength vanishes are discussed separately in the
    text.
    }
    \label{fig:two_particle_phase_diagrams}
\end{figure}

Figure~\ref{fig:two_particle_phase_diagrams} shows both the magnitude
of the critical coupling and the nature of the corresponding
instability. The Hopf instability, as shown in orange, occurs first over much of the
interior of the parameter plane. Therefore, a finite-frequency mode provides
the predominant linear route away from the stationary antipodal
state. This conclusion concerns only the linear onset of activity. A
pitchfork bifurcation may produce a uniformly translating pair with
fixed separation, whereas a Hopf bifurcation may produce periodic
oscillations of the particle velocities and relative separation.
The stability of these nonlinear states, and the possible emergence
of drifting, slipping, or more complicated motion farther from onset,
must be subsequently determined from direct numerical simulations.

The pitchfork-first regions, shown in blue, form two lobes related by particle
exchange. They occur principally when chemical production and
phoretic response are concentrated on different particles. For
example, at small \(\theta\) and large \(\phi\), particle 1 is the
stronger producer, $s_1>s_2,$ whereas particle 2 is the stronger responder, $  \lambda_2>\lambda_1.$
The second lobe represents the equivalent configuration with the
particle labels exchanged. The zero-frequency mode is therefore
favoured when the producer and responder roles are distributed
between the two particles, although such opposed asymmetries are
not a strict requirement: the lobes also reach the equal-response
line \(\phi=\pi/4\) near the source-strength endpoints. By contrast, 
Hopf onset generally dominates when the same particle
is both the stronger producer and the stronger responder, as well
as throughout much of the neighbourhood of the identical-particle
point $    (\theta,\phi)  =  \left(\frac{\pi}{4},\frac{\pi}{4}\right).$
The stability diagram thus relates the form of the first instability
to how chemical production and phoretic response are distributed
between the particles.

A useful interpretation 
here is that opposing production and response
asymmetries make the chemical coupling strongly non-reciprocal:
the stronger producer drives the stronger responder, while the
reverse coupling is weak. This directional coupling favours a
quasistatic co-moving response and hence the zero-frequency mode. 
However, this favouring of the zero-frequency mode relative to the Hopf mode does not
necessarily reduce the absolute instability threshold. Indeed,
the threshold diverges on approaching the maximally anti-aligned corners $    (\theta,\phi)
    =
    \left(0,\frac{\pi}{2}\right)
    \ \text{and}\
    \left(\frac{\pi}{2},0\right).$ At the first corner,
$$
    s_1=s,\qquad s_2=0,
    \qquad
    \lambda_1=0,\qquad\lambda_2=\lambda,
$$
so particle 1 acts as a fixed source and particle 2 as a
non-emitting responder. Because the responder cannot modify either
the field or the source motion, the feedback loop required to
destabilise the antipodal state is broken. The divergence therefore
reflects the loss of dynamical feedback, rather than the absence of
a chemical gradient. Divergence at the opposite corner follows similarly by exchanging the
particle labels.

The edges \(\theta=0\) and \(\theta=\pi/2\), away from their
endpoints, reduce to the active-source--tracer problem considered
above. Because the tracer does not modify the chemical field, the
source dynamics decouple from the tracer and can lose stability only
through the single-particle pitchfork instability. For a tracer with
nonzero positive mobility, the locking threshold tends to zero as
the source velocity tends to zero. The tracer therefore locks to the
source immediately above onset. The maximally aligned corners $ (\theta,\phi)=(0,0)
    \ \text{and}\ (\theta,\phi)=\left(\frac{\pi}{2},\frac{\pi}{2}\right)$
instead reduce to a single active particle accompanied by a particle
that neither produces nor responds to the chemical. Their onset
thresholds must consequently recover the single-particle result.

Particle exchange provides a useful consistency check. From the
angular parametrisation \eqref{eq:angular_parametrisation},
interchanging particles \(1\leftrightarrow2\) is equivalent to
\begin{equation}
    (\theta,\phi)
    \longrightarrow
    \left(
        \frac{\pi}{2}-\theta,
        \frac{\pi}{2}-\phi
    \right).
\end{equation}
Indeed, this transformation exchanges
\(\cos\theta\leftrightarrow\sin\theta\) and
\(\cos\phi\leftrightarrow\sin\phi\), and hence exchanges
\((s_1,\lambda_1)\) with \((s_2,\lambda_2)\).
Since relabelling the particles cannot change the physical
instability threshold, the stability diagram must be invariant
under this transformation. Geometrically, it corresponds to a
rotation through \(\pi\) about \((\frac{\pi}{4}, \frac{\pi}{4})\), exchanging the
two pitchfork lobes and the two high-threshold corners.

\begin{figure}[htbp]
    \centering

    \begin{subfigure}[t]{0.48\textwidth}
        \centering
        \includegraphics[width=\linewidth]{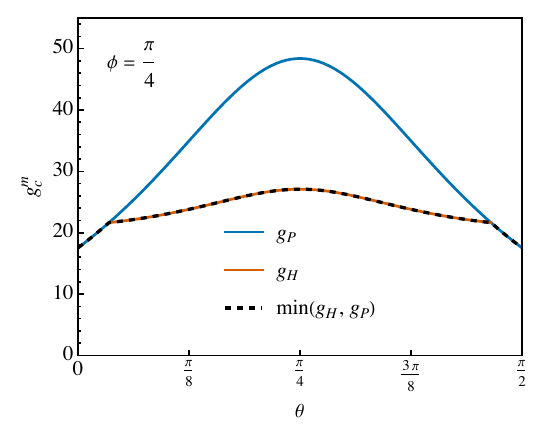}
        \caption{
        Variation with \(\theta\) at fixed \(\phi=\pi/4\).
        }
        \label{fig:cross_section_fixed_phi_pi4}
    \end{subfigure}
    \hfill
    \begin{subfigure}[t]{0.48\textwidth}
        \centering
        \includegraphics[width=\linewidth]{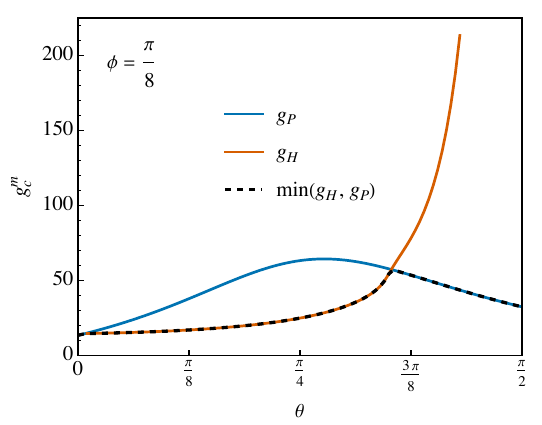}
        \caption{
        Variation with \(\theta\) at fixed \(\phi=\pi/8\).
        }
        \label{fig:cross_section_fixed_phi_pi8}
    \end{subfigure}

    \vspace{0.6cm}

    \begin{subfigure}[t]{0.48\textwidth}
        \centering
        \includegraphics[width=\linewidth]{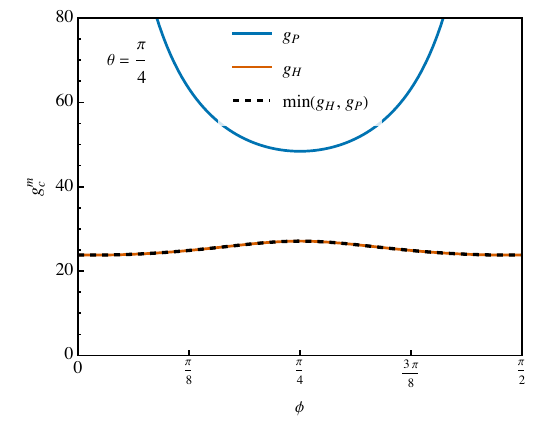}
        \caption{
        Variation with \(\phi\) at fixed \(\theta=\pi/4\).
        }
        \label{fig:cross_section_fixed_theta_pi4}
    \end{subfigure}
    \hfill
    \begin{subfigure}[t]{0.48\textwidth}
        \centering
        \includegraphics[width=\linewidth]{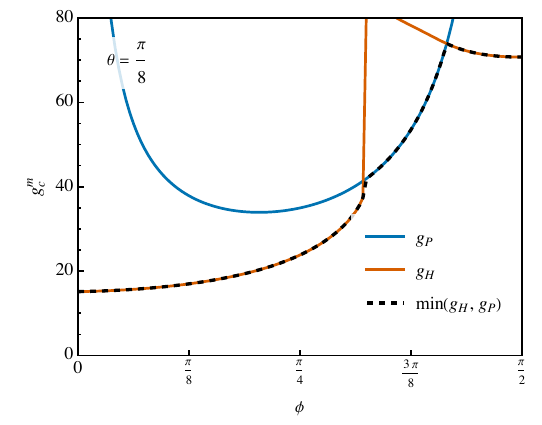}
        \caption{
        Variation with \(\phi\) at fixed \(\theta=\pi/8\).
        }
        \label{fig:cross_section_fixed_theta_pi8}
    \end{subfigure}

    \caption{
    Representative cross-sections of the two-particle stability
    diagram. Blue and orange curves show \(g_{\mathrm P}\) and
    \(g_{\mathrm H}\), respectively, while the dashed black curve
    gives the physical threshold
    \(g_c^{m}=\min\{g_{\mathrm P},g_{\mathrm H}\}\).
    The left-hand panels contain one balanced parameter,
    \(\theta=\pi/4\) or \(\phi=\pi/4\), whereas the right-hand
    panels illustrate representative asymmetric configurations.
    }
    \label{fig:stability_cross_sections}
\end{figure}

The cross-sections in Fig.~\ref{fig:stability_cross_sections} make
the competition between the two instabilities more explicit. For
equal response coefficients, \(\phi=\pi/4\), particle exchange maps
\(\theta\) to \(\pi/2-\theta\), and the threshold is therefore
symmetric about \(\theta=\pi/4\). Hopf onset dominates around the
identical-particle point, whereas pitchfork onset occurs near the
source-strength endpoints. Along this cut, the first-instability
threshold reaches its maximum for identical particles. For the
asymmetric choice \(\phi=\pi/8\), this reflection symmetry
is lost. As \(\theta\) increases, the lower envelope initially
follows the rising Hopf threshold and then switches to the decreasing
pitchfork threshold. Their intersection produces the pronounced
maximum in the physical threshold.

The mobility scans in Fig.~\ref{fig:cross_section_fixed_theta_pi4} and \ref{fig:cross_section_fixed_theta_pi8} demonstrate that source and response asymmetries
are not interchangeable. For equal source strengths,
\(\theta=\pi/4\), Hopf onset remains dominant throughout the
entire interval, even though the pitchfork threshold varies
strongly with \(\phi\). For \(\theta=\pi/8\), the lower envelope
instead passes through Hopf-, pitchfork-, and Hopf-first regions as
\(\phi\) is increased. For the reader's clarity, this sequence describes the type of first
instability obtained at different fixed parameter pairs; it does
not represent three successive bifurcations during a single sweep
of the coupling \(g\).

\subsubsection{Comparison of activity threshold with direct numerical simulations}

\begin{figure}[htbp]
    \centering

    \begin{subfigure}[t]{0.48\textwidth}
        \centering
        \includegraphics[width=\linewidth]{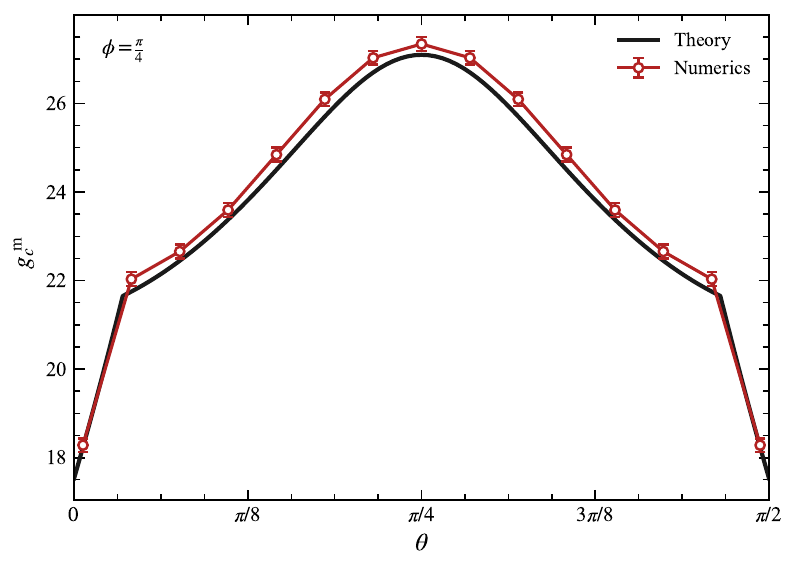}
        \caption{\(\phi=\pi/4\).}
        \label{fig:transition-phi-pi4}
    \end{subfigure}
    \hfill
    \begin{subfigure}[t]{0.48\textwidth}
        \centering
        \includegraphics[width=\linewidth]
        {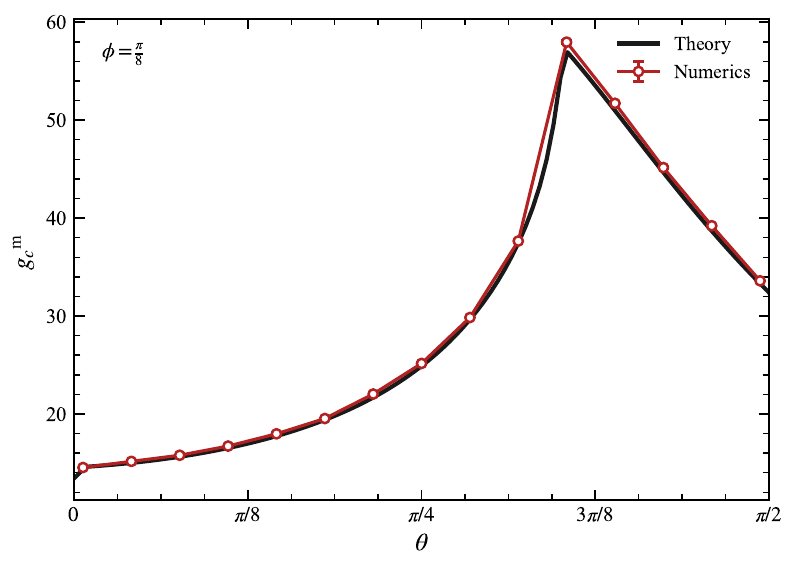}
        \caption{\(\phi=\pi/8\).}
        \label{fig:transition-phi-pi8}
    \end{subfigure}

    \caption{
    Comparison between the theoretical and numerical
    inactive--active boundaries as functions of the source-asymmetry
    angle \(\theta\). The black curves show the theoretical
    first-instability threshold
    \(g_c^m=\min\{g_{\mathrm P},g_{\mathrm H}\}\), while the red
    markers show the boundary obtained from direct simulations.
    The error bars span the final interval between the largest
    coupling classified as inactive and the smallest classified
    as active.
    }
    \label{fig:transition-theory-numerics}
\end{figure}

We test the linear stability predictions by direct numerical
integration of the coupled particle--field equations using the
IMEX Euler method described in Appendix~\ref{sec:numerics}. For each
parameter pair \((\theta,\phi)\), the late-time activity was
measured through the root-mean-square particle velocity
\begin{equation}
    v_{\mathrm{RMS}}
    =
    \left[
        \frac{1}{2T}
        \int_t^{t+T}
        \left(
            \dot X_1(t')^2+\dot X_2(t')^2
        \right)
        \,\mathrm{d}t'
    \right]^{1/2}.
    \label{eq:two_particle_rms_velocity}
\end{equation}
Unlike the mean signed velocity, this quantity detects both steady
translation and oscillatory motion. A trajectory was classified as
inactive when \(v_{\mathrm{RMS}}\) decayed below the prescribed
tolerance, and as active when motion above that tolerance persisted
over successive observation windows. After an initial coarse scan 
in \(g\), the transition was refined
by bisection. Each numerical point in
Fig.~\ref{fig:transition-theory-numerics} is the midpoint of the
final inactive--active bracket, while its error bar gives the width
of that bracket. These error bars therefore measure the resolution
of the coupling scan rather than the total numerical uncertainty.

The simulations employed the Fej\'er-regularised Fourier method
introduced in Appendix~\ref{subsec:fourier_regularisation}, with
\(M=512\) and \(\Delta t=2\times10^{-4}\). The initial antipodal
state was perturbed by a relative displacement of \(10^{-4}\).
The RMS velocity was evaluated over windows of duration \(T=8\)
after a minimum integration time of \(32\). Simulations for which
the classification remained ambiguous were extended to a maximum
time of \(160\). The activity tolerance was \(10^{-5}\), and the
final coupling brackets had a maximum width \(\Delta g=0.5\).

The simulations reproduce both the symmetric boundary at
\(\phi=\pi/4\) and the strongly asymmetric boundary at
\(\phi=\pi/8\). In particular, they capture the pronounced maximum
in the latter case, where the first instability changes from Hopf
to pitchfork type. The numerical thresholds lie slightly above the
theoretical curves over much of the sampled range, with the clearest
systematic offset near the centre of the equal-response cut.
This positive offset is consistent with two finite-resolution
effects. First, Fej\'er regularisation smooths each point source at
finite \(M\) and can therefore shift the instability threshold.
Second, close to onset, the small growth rate produces a long
transient, so a weakly unstable trajectory may be classified as
inactive over a finite observation time. Time discretisation may
also contribute to the remaining difference. Nevertheless, the
close agreement across both parameter cuts supports the predicted
inactive--active boundary. Because the classification uses only the
magnitude of the particle velocities, it verifies the onset of
motion but does not independently distinguish pitchfork from Hopf
instability. We now proceed by examining the nonlinear dynamics of the active regime via direct
numerical simulations.

\subsubsection{Nonlinear two-particle dynamics and comparison with camphor-disk experiments}
\label{sec:two_particle_nonlinear_dynamics}

We first examine representative asymmetric configurations lying well
inside the Hopf-first and pitchfork-first regions, and then consider
the activity-dependent dynamics of two identical particles. To
distinguish collective translation from changes in the interparticle
separation, we introduce the relative velocity
\begin{equation}
    U_{\mathrm{rel}}
    =
    U_2-U_1,
    \label{eq:two_particle_velocity_coordinates}
\end{equation}
together with the oriented separation
\begin{equation}
    \Delta(t)
    =
    \bigl[X_2(t)-X_1(t)\bigr]\bmod L,
    \qquad
    0\leq\Delta<L.
    \label{eq:two_particle_relative_separation}
\end{equation}
A uniformly translating locked state has
\(U_{\mathrm{rel}}=0\) and constant \(\Delta\), whereas periodic
variations of \(U_{\mathrm{rel}}\) and \(\Delta\) indicate
oscillatory relative motion. Unwrapped particle positions are used
to determine the net circulation around the periodic domain.

\begin{figure}[htbp]
    \centering

    \begin{subfigure}[t]{0.48\textwidth}
        \centering
        \includegraphics[width=\linewidth]{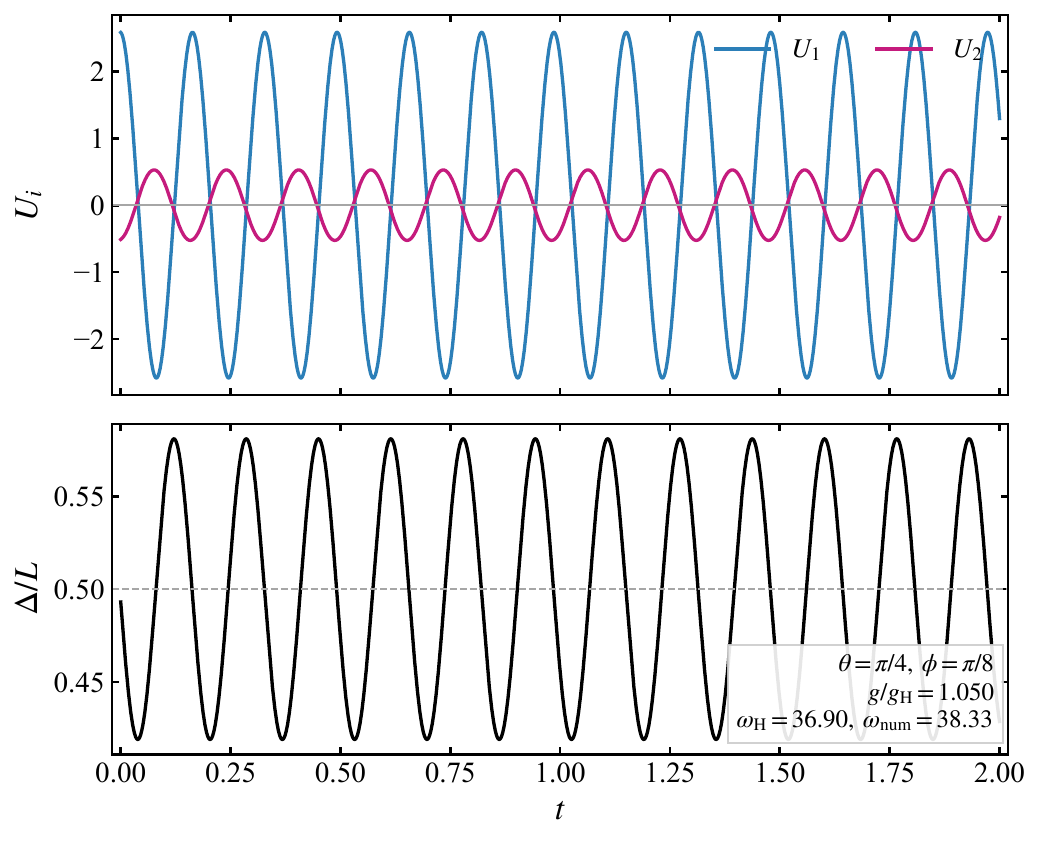}
        \caption{
        Hopf-first case:
        \(\theta=\pi/4\), \(\phi=\pi/8\), and
        \(g=1.05g_{\mathrm H}\).
        }
        \label{fig:two_particle_near_onset_hopf}
    \end{subfigure}
    \hfill
    \begin{subfigure}[t]{0.48\textwidth}
        \centering
        \includegraphics[width=\linewidth]{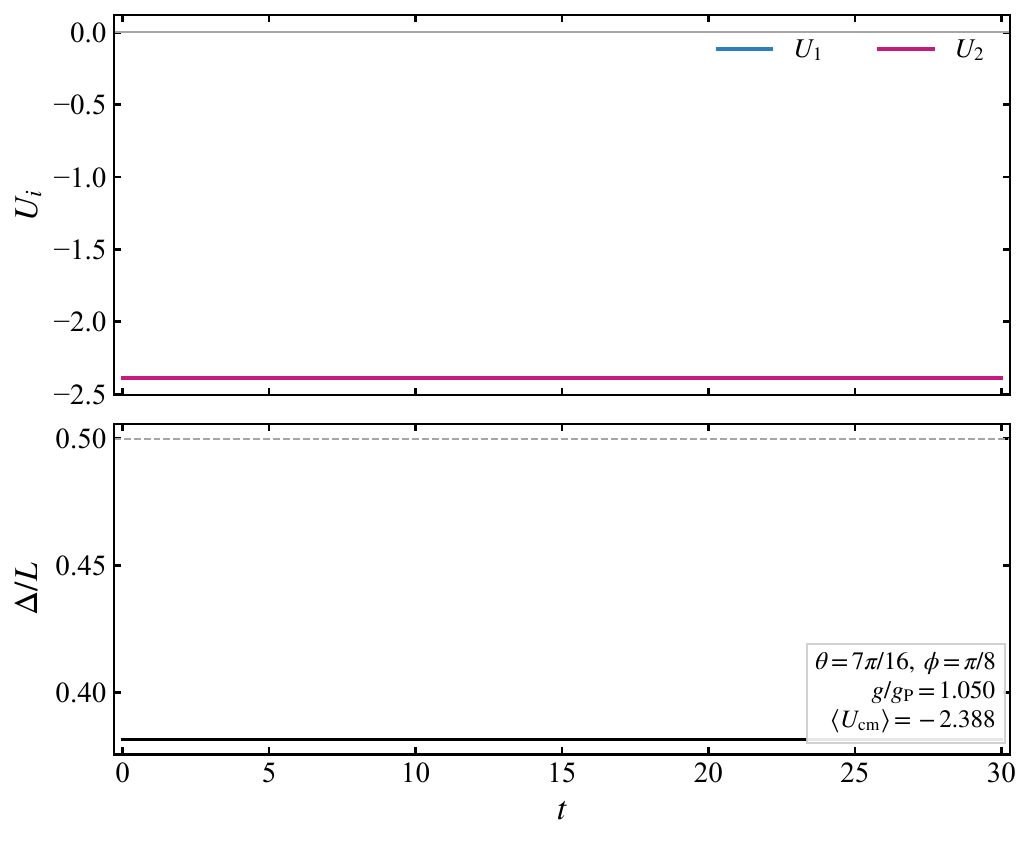}
        \caption{
        Pitchfork-first case:
        \(\theta=7\pi/16\), \(\phi=\pi/8\), and
        \(g=1.05g_{\mathrm P}\).
        }
        \label{fig:two_particle_near_onset_pitchfork}
    \end{subfigure}

    \caption{
    Nonlinear dynamics immediately above the first instability of
    the stationary antipodal state. Each panel shows the particle
    velocities \(U_1\) and \(U_2\), together with the relative
    separation \(\Delta\), after the initial transient.
    (a) Above the Hopf threshold, the velocities and separation
    undergo sustained periodic oscillations.
    (b) Above the pitchfork threshold, the particles approach a
    uniformly translating locked state with common velocity
    \(U_\star\) and fixed separation \(\Delta_\star\).
    }
    \label{fig:two_particle_near_onset}
\end{figure}

For $    (\theta,\phi)
    =
    \left(\frac{\pi}{4},\frac{\pi}{8}\right),$ the first instability is Hopf-type, so that
\(g_c^m=g_{\mathrm H}\). At \(g=1.05g_{\mathrm H}\), an initial
perturbation develops into sustained oscillations of the particle
velocities and relative separation
[Fig.~\ref{fig:two_particle_near_onset}(a)]. Their persistence after
the initial transient distinguishes the nonlinear state from damped
relaxation towards the antipodal configuration. The oscillation frequency is determined from the dominant peak in
the late-time Fourier spectrum of \(U_{\mathrm{rel}}\). At
\(g/g_{\mathrm H}=1.05\), the measured value is
\(\omega_{\mathrm{num}}=\omega(g)=38.327\), compared with the predicted linear onset frequency
\(\omega_{\mathrm H}=\omega(g_H)=36.90\). The relative difference is
\begin{equation}
    \frac{
        \left|\omega_{\mathrm{num}}-\omega_{\mathrm H}\right|
    }{
        \omega_{\mathrm H}
    }
    =0.0387.
\end{equation}
This agreement supports the identification of the observed periodic
state with the Hopf mode predicted by the linear stability analysis.
The remaining difference is consistent with the finite distance
above onset and the finite Fourier cutoff used in the simulation.

For comparison, the configuration $    (\theta,\phi)
    =
    \left(\frac{7\pi}{16},\frac{\pi}{8}\right)$
lies well inside the pitchfork-first region, where
\(g_c^m=g_{\mathrm P}\). At \(g=1.05g_{\mathrm P}\), the two
velocities approach a common nonzero value while the separation
becomes constant:
$$
    U_1(t),U_2(t)\longrightarrow U_\star,
    \qquad
    \Delta(t)\longrightarrow\Delta_\star.
$$
The resulting state is therefore a uniformly translating locked pair
[Fig.~\ref{fig:two_particle_near_onset}(b)]. Reflection symmetry
produces an equivalent branch travelling in the opposite direction,
and the branch selected in a given simulation is determined by the
initial perturbation.

Supplementary Videos~S1 and S2 show the corresponding evolution of
the particles and chemical field. Video~S1 illustrates the periodic
deformation and relaxation of the field during the Hopf oscillation,
whereas Video~S2 shows the stationary profile established in the
frame translating with the locked pair. Additional simulations near
the boundary between the Hopf-first and pitchfork-first regions
approached either periodic relative motion or uniform translation.
No persistent state combining net rotation with relative oscillation
was identified numerically. Thus, within the parameter range examined, the
crossover separates the two onset mechanisms but does not generate
an additional long-time dynamical state.

We finally consider two identical particles, for which
\begin{equation}
    \theta=\phi=\frac{\pi}{4},
    \qquad
    s_1=s_2=\frac{s}{\sqrt{2}},
    \qquad
    \lambda_1=\lambda_2=\frac{\lambda}{\sqrt{2}}.
\end{equation}
The antipodal state first loses stability through a Hopf bifurcation,
which develops nonlinearly into a symmetric reversal oscillation:
the particles move with equal and opposite velocities while the mean
centre-of-mass velocity remains zero. Increasing the activity raises
both the oscillation amplitude and frequency, but produces no
secondary transition over the range
\(1<g/g_c^m\leq5\). Simulations initiated from different particle
separations and chemical-field perturbations also converged to the
same reversal-oscillation state. The oscillatory dynamics are
therefore robust throughout the activity range investigated in the
vicinity of the identical-particle configuration.

This reversal oscillation is qualitatively consistent with one of
the dynamical modes reported for two symmetric camphor disks in an annular water
channel by Nishi \textit{et al.}~\cite{Nishi2015}. Their experiments
also exhibited stop-and-move rotation, equally spaced rotation, and
clustered rotation. Real camphor disks, however, cannot be expected
to have exactly equal dissolution rates, shapes, or phoretic
responses. The present stability diagram shows that the onset
dynamics depend sensitively on how the source strengths and
phoretic mobilities are distributed between the particles. A
sufficient change in these relative properties can move the system
from a Hopf-first region, producing reversal oscillations, into a
pitchfork-first region, producing persistent rotation of a locked
pair. This latter state may then be further classified as either equally spaced rotation or clustered rotation, depending on the steady-state value $\Delta_\star$. Particle-to-particle variability may therefore contribute to
the diversity of modes observed experimentally, even when the disks
are nominally identical.

This interpretation should remain qualitative. Close to the
identical-particle point \((\theta,\phi)=(\pi/4,\pi/4)\), the Hopf
mode remains dominant at onset, and the subsequent activity sweep
detects only reversal oscillations. In particular, the present
overdamped point-particle model does not produce a steady
stop-and-move rotational state. Reproducing the full range of modes
reported by Nishi \textit{et al.} may require additional effects,
such as finite particle size, mechanical inertia, hydrodynamic
coupling, or more detailed interfacial dynamics. Nevertheless, the
results demonstrate that mutual coupling through
a slowly relaxing chemical field can coordinate the motion of two
individually isotropic particles. The recovery of analogous
reversal and clustered-rotation states indicates that chemical
memory and periodic geometry alone are sufficient to generate
several of the characteristic collective motions observed in the
camphor-disk system.

\section{Conclusion}
\label{sec:discussion_conclusion}

In this work, we have developed a minimal framework for
self-phoretic motion and chemically mediated interactions between particles on a
circular domain. Particles release a chemical that
diffuses around and evaporates from  the ring, while their velocities are
determined by the local concentration gradient. Despite its
simplicity, the model captures the delayed feedback between particle
motion and the chemical field that underlies both spontaneous
self-propulsion and collective dynamics. The periodic geometry also
makes it possible to determine explicitly how finite system size and
chemical persistence modify these behaviours.

For a single particle, the stationary symmetric state loses
stability through a pitchfork bifurcation, producing two equivalent
states that travel in opposite directions around the ring. The
nonlinear travelling solutions were obtained analytically and shown
to be stable both close to onset and over the broader activity range
examined numerically. The critical activity depends on the ratio
between the chemical decay length and the domain size, demonstrating
that periodic confinement can substantially modify the onset of
self-propulsion. In the large-domain or short-decay-length limit, the
periodic result approaches the corresponding threshold on an
unbounded domain. Direct simulations reproduce the predicted
transition, travelling velocities, and convergence towards the
continuum limit.

The source--tracer problem isolates the effect of a travelling
chemical profile on a second, non-producing particle. Depending on
the tracer mobility, the tracer either locks at a fixed separation
from the active source or continually slips relative to it and
circulates around the ring. The locking condition and its stability
were obtained analytically, yielding a phase boundary in close
agreement with the numerical simulations. This intermediate problem
illustrates how a self-generated field can entrain another particle
even in the absence of reciprocal coupling, while also identifying
the conditions under which the tracer cannot attain the source
velocity.

For two chemically active particles, reciprocal coupling introduces
a second instability mechanism. The stationary antipodal state may
lose stability either through a zero-frequency pitchfork mode,
leading to a uniformly translating locked pair, or through a
finite-frequency Hopf mode, leading to oscillatory relative motion.
Which instability occurs first depends sensitively on the relative
source strengths and phoretic mobilities of the two particles. This
dependence reflects the structure of the chemical feedback loop:
redistributing the producer and responder roles can favour one mode
over the other, whereas eliminating one direction of the feedback
causes the corresponding instability threshold to diverge. The predicted transition boundaries are supported by direct
simulations along representative parameter cuts,
and representative nonlinear trajectories confirm the emergence of
uniform translation and sustained oscillation above the pitchfork
and Hopf thresholds, respectively.

For identical particles, the Hopf instability develops into a robust
reversal oscillation with no mean centre-of-mass motion. This state
persists throughout the activity range investigated and is
insensitive to the initial conditions considered. It qualitatively
recovers the reversal oscillations observed for pairs of camphor
disks in annular channels by Nishi \textit{et al.}~\cite{Nishi2015}.
The strong dependence of the predicted dynamics on the relative
chemical production rates and mobilities also suggests that
particle-to-particle variability may contribute to the broader range
of rotational and oscillatory states observed experimentally.
Nevertheless, the present model does not produce steady
stop-and-move rotation, and the comparison should therefore remain
qualitative.

Several physical effects have deliberately been omitted, including
finite particle size, excluded-volume interactions, mechanical
inertia, hydrodynamic coupling, fluctuations, and nonlinear or
time-dependent chemical release. Such effects may become important
when particles approach one another closely or when more complicated
rotational states arise. Extending the model to include these
features, or to larger particle populations, would provide a natural
route towards connecting the analytically tractable two-particle
problem with clustering, density waves, and collective transport in
experiments.

Overall, the results show that spontaneous propulsion, entrainment,
reversal oscillations, and collective rotation can all arise from
the same basic mechanism: delayed coupling through a self-produced
chemical field. The selected motion is controlled not only by the
overall activity, but also by periodic confinement and by how
chemical production and response are distributed among the
particles. These ingredients therefore provide a compact framework
for understanding how individual self-phoretic motion develops into
distinct forms of chemically mediated collective dynamics.

\appendix
\section{Numerical methods}
\label{sec:numerics}

\subsection{Semi-implicit Fourier time-stepping scheme}

We integrate the coupled particle--field equations using a truncated
Fourier representation and a first-order semi-implicit scheme.
Lengths and times are scaled by $L$ and $L^2/D$, respectively,
while a reference source strength $s_0>0$ sets the concentration
scale:
\begin{equation}
    x^*=\frac{x}{L},
    \qquad
    t^*=\frac{Dt}{L^2},
    \qquad
    c^*=\frac{D}{s_0L}c.
\end{equation}
The corresponding dimensionless parameters are
\begin{equation}
    \mu^*=\frac{\mu L^2}{D},
    \qquad
    \lambda_i^*=\frac{\lambda_i s_0L}{D^2},
    \qquad
    s_j^*=\frac{s_j}{s_0}.
    \label{eq:numerical_dimensionless_parameters}
\end{equation}
For a single particle, we choose $s_0=s$, giving $s^*=1$ and
$\lambda^*=\lambda sL/D^2$. Since the spatially uniform concentration does not affect
particle motion, only the nonzero Fourier modes need to be
evolved. On the unit periodic domain, we approximate the
spatially varying field by
\begin{equation}
    c^*(x^*,t^*)-\overline{c^*}(t^*)
    \simeq
    \sum_{\substack{n=-M\\n\neq0}}^{M}
    a_n(t^*)\ee^{\ii q_nx^*},
    \qquad
    q_n=2\pi n,
    \label{eq:numerical_fourier_expansion}
\end{equation}
where $\overline{c^*}$ is the spatial mean and $M$ is the
largest retained positive mode number. Here, $a_n$ denotes
the dimensionless Fourier amplitude. The mode equations are
\begin{equation}
    \frac{\mathrm{d}a_n}{\mathrm{d}t^*}
    =
    -\epsilon_n^*a_n
    +
    \sum_{j=1}^{N}
    s_j^*\ee^{-\ii q_nX_j^*(t^*)},
    \qquad
    \epsilon_n^*=q_n^2+\mu^*.
    \label{eq:numerical_mode_dynamics}
\end{equation}
The dimensionless particle velocity,
\begin{equation}
    U_i=\frac{L}{D}V_i
       =\frac{\mathrm{d}X_i^*}{\mathrm{d}t^*},
\end{equation}
is evaluated as
\begin{equation}
    U_i(t^*)
    =
    -\lambda_i^*
    \operatorname{Re}
    \left[
        \sum_{\substack{n=-M\\n\neq0}}^{M}
        \ii q_na_n(t^*)\ee^{\ii q_nX_i^*(t^*)}
    \right].
    \label{eq:numerical_velocity}
\end{equation}
Conjugate symmetry, $a_{-n}=\overline{a_n}$, makes this sum
real; taking its real part explicitly removes residual
imaginary contributions from floating-point error.

We let $a_n^m$ and $X_i^m$ denote the Fourier amplitudes and
dimensionless particle positions at
$t_m^*=m\Delta t^*$. Each timestep consists of three updates.
First, the particle velocities are evaluated from the current
field and positions:
\begin{equation}
    U_i^m
    =
    -\lambda_i^*
    \operatorname{Re}
    \left[
        \sum_{\substack{n=-M\\n\neq0}}^{M}
        \ii q_na_n^m\ee^{\ii q_nX_i^m}
    \right].
    \label{eq:explicit_velocity_update}
\end{equation}
The positions are then advanced by a forward Euler step,
\begin{equation}
    X_i^{m+1}
    =
    X_i^m+\Delta t^*U_i^m.
    \label{eq:explicit_position_update}
\end{equation}
Finally, the field is updated using the new source positions:
\begin{equation}
    a_n^{m+1}
    =
    \frac{
        a_n^m
        +
        \Delta t^*
        \displaystyle\sum_{j=1}^{N}
        s_j^*\ee^{-\ii q_nX_j^{m+1}}
    }{
        1+\Delta t^*\epsilon_n^*
    }.
    \label{eq:semiimplicit_field_update}
\end{equation}
Unwrapped particle positions are retained to record displacement
around the domain, while positions reduced modulo unity are
used to evaluate the periodic Fourier phases. We note that diffusion and evaporation are treated implicitly through Eq.~\eqref{eq:semiimplicit_field_update}, whereas the source
term is evaluated from the explicitly updated positions.
This avoids the diffusive timestep restriction of a fully
explicit field update, which becomes increasingly severe as
$\epsilon_n^*\sim n^2$ at large mode number. It does not remove
the need to resolve the coupled particle--field dynamics:
the overall scheme remains first-order accurate in
$\Delta t^*$, and convergence must be checked by reducing
the timestep and increasing $M$, particularly at large
mobilities.
\subsection{Initial conditions and Fourier regularisation}
\label{subsec:fourier_regularisation}

The initial chemical field is chosen as the stationary profile
generated by particles held at their initial positions:
\begin{equation}
    a_n^0
    =
    \frac{1}{\epsilon_n^*}
    \sum_{j=1}^{N}
    s_j^*\ee^{-\ii q_nX_j^0},
    \qquad n\neq0.
    \label{eq:initial_field_modes}
\end{equation}
For a single particle, we take $X^0=0$ and $s^*=1$, so that
$a_n^0=1/\epsilon_n^*$.
A small dimensionless velocity perturbation $U_{\mathrm{seed}}$
is applied during the first timestep to break reflection symmetry
and select a direction of motion. Subsequent evolution follows
the coupled particle--field equations without external forcing.
For comparison with the analytical velocity curves, we report
\begin{equation}
    \widehat v_i
    =
    \frac{LV_i}{2D}
    =
    \frac{U_i}{2}.
\end{equation}

For the standard two-particle simulations, we initialise the
positions close to the antipodal configuration,
\begin{equation}
    X_1^0=0,
    \qquad
    X_2^0=\frac{1}{2}+\eta,
    \qquad
    \eta=10^{-2},
\end{equation}
and construct the corresponding field using
Eq.~\eqref{eq:initial_field_modes}.
The displacement $\eta$ perturbs the relative separation away
from the reflection-symmetric antipodal state.
The scans used to locate the two-source instability threshold
employ smaller positional perturbations, and also require care with Fourier truncation.
The concentration gradient is discontinuous at each point source,
so a sharp cutoff can introduce oscillatory errors into the
computed particle velocities. At the antipodal separation, the
Fourier phase entering the interaction between particles is
\begin{equation}
    \ee^{\ii q_n(X_i-X_j)}
    =
    \ee^{\pm\ii q_n/2}
    =
    (-1)^n.
\end{equation}
The interaction sums therefore contain alternating contributions,
and their truncation errors can depend on whether the cutoff $M$
is even or odd. To reduce this sensitivity, we apply Fej\'er
weights to the source terms in the two-source threshold scans:
\begin{equation}
    w_n=1-\frac{|n|}{M+1},
    \qquad |n|\leq M.
    \label{eq:fejer_weights}
\end{equation}
These weights decrease linearly with mode number, suppressing
the modes nearest the cutoff. The field update then becomes
\begin{equation}
    a_n^{m+1}
    =
    \frac{
        a_n^m
        +
        \Delta t^* w_n
        \displaystyle\sum_{j=1}^{N}
        s_j^*\ee^{-\ii q_nX_j^{m+1}}
    }{
        1+\Delta t^*\epsilon_n^*
    },
    \label{eq:fejer_field_update}
\end{equation}
with the corresponding stationary initial field
\begin{equation}
    a_n^0
    =
    \frac{w_n}{\epsilon_n^*}
    \sum_{j=1}^{N}
    s_j^*\ee^{-\ii q_nX_j^0},
    \qquad n\neq0.
    \label{eq:fejer_initial_field}
\end{equation}
Since the weights are already incorporated into the evolved
amplitudes, the velocity is evaluated using
Eq.~\eqref{eq:numerical_velocity}, without an additional factor
of $w_n$.

At finite $M$, this procedure replaces each point source by a
spatially smoothed periodic source and can therefore shift the
measured instability threshold. As $M\to\infty$, the regularised
source approaches the periodic delta function, with $w_n\to1$
for every fixed mode. Convergence with respect to $M$ must
therefore be assessed alongside timestep and observation-time
convergence. The Fourier cutoff, timestep, perturbation amplitudes,
and observation times are specified with each set of numerical
results.

\section{Monotonicity of the concentration gradient}
\label{app:active_tracer_monotonicity}

Here we show that the concentration gradient generated by a uniformly
moving point source is monotonic between successive images of the source.
In the reference frame moving at velocity $v$, the stationary concentration
profile satisfies
\begin{equation}
    D c_v''(x)+v c_v'(x)-\mu c_v(x)
    =
    -s\delta_L(x),
    \label{eq:appendix_comoving_field}
\end{equation}
with periodic boundary conditions on a domain of length $L$. Away from the
source, i.e.\ for $0<x<L$, this reduces to
\begin{equation}
    D c_v''(x)+v c_v'(x)-\mu c_v(x)=0.
    \label{eq:appendix_comoving_homogeneous}
\end{equation}
The corresponding characteristic roots are
\begin{equation}
    r_\pm
    =
    \frac{-v\pm\sqrt{v^2+4D\mu}}{2D}.
    \label{eq:appendix_characteristic_roots}
\end{equation}
For $D>0$ and $\mu>0$, these satisfy
\begin{equation}
    r_-<0<r_+.
\end{equation}

The periodic solution of Eq.~\eqref{eq:appendix_comoving_field}, with the
source located at $x=0$, can be written on the interval $0<x<L$ as
\begin{equation}
    c_v(x)
    =
    \frac{s}{D(r_+-r_-)}
    \left[
        \frac{e^{r_-x}}{1-e^{r_-L}}
        +
        \frac{e^{r_+(x-L)}}{1-e^{-r_+L}}
    \right].
    \label{eq:appendix_periodic_concentration}
\end{equation}
This expression is continuous across the source and satisfies the gradient
jump condition
\begin{equation}
    c_v'(0^+)-c_v'(L^-)
    =
    -\frac{s}{D}.
\end{equation}

Differentiating Eq.~\eqref{eq:appendix_periodic_concentration} twice gives
\begin{equation}
    c_v''(x)
    =
    \frac{s}{D(r_+-r_-)}
    \left[
        \frac{r_-^2e^{r_-x}}{1-e^{r_-L}}
        +
        \frac{r_+^2e^{r_+(x-L)}}{1-e^{-r_+L}}
    \right].
    \label{eq:appendix_concentration_second_derivative}
\end{equation}
Since $r_+-r_->0$ and
\begin{equation}
    1-e^{r_-L}>0,
    \qquad
    1-e^{-r_+L}>0,
\end{equation}
all factors inside the square brackets in
Eq.~\eqref{eq:appendix_concentration_second_derivative} are strictly
positive for $0<x<L$. It follows that
\begin{equation}
    \operatorname{sgn}\!\left[c_v''(x)\right]
    =
    \operatorname{sgn}(s),
    \qquad 0<x<L.
\end{equation}
Thus, for a positive source strength,
\begin{equation}
    c_v''(x)>0,
    \qquad 0<x<L,
\end{equation}
so that $c_v'(x)$ is strictly increasing between successive images of the
source. Consequently, the range of $c_v'$ on $(0,L)$ is bounded by its one-sided
values at the source. The active tracer therefore admits a unique interior
locked separation if and only if
\begin{equation}
    -\frac{v}{\lambda_2}
    \in
    \left(
        \min\!\left\{c_v'(0^+),c_v'(L^-)\right\},
        \max\!\left\{c_v'(0^+),c_v'(L^-)\right\}
    \right).
\end{equation}
Using the one-sided gradients derived in
Eqs.~\eqref{eq:active_tracer_cusp_plus} and
\eqref{eq:active_tracer_cusp_minus}, this is equivalent to
\begin{equation}
    \left|
        v\left(
            \frac{1}{\lambda_1}
            -
            \frac{1}{\lambda_2}
        \right)
    \right|
    <
    \frac{|s|}{2D},
\end{equation}
confirming that Eq.~\eqref{eq:active_tracer_existence_condition} is both
necessary and sufficient for the existence of an interior locked solution.

\section{Derivation of the stationary two-source response determinant}
\label{app:two_source_response_determinant}

Here we provide the intermediate steps leading to
Eq.~\eqref{eq:two_source_stationary_determinant}. We consider the stationary
antipodal state
\begin{equation}
v=0,
\qquad
d_{12}=\frac{1}{2},
\end{equation}
in the dimensionless units $L=D=1$. For convenience, define
\begin{equation}
\alpha=\sqrt{\mu},
\qquad
\beta=\sqrt{\mu+\ii\omega},
\end{equation}
where the principal branch of the square root is understood. For a stationary source on the periodic domain, the regularised self-kernel
and the kernel evaluated at the antipodal point are
\begin{align}
    K_z^{(2)}(0,0;0)
    &=
    -\frac{\sqrt{z}}{2}
    \coth\left(\frac{\sqrt{z}}{2}\right),
    \label{eq:app_two_source_self_kernel}
    \\[1ex]
    K_z^{(2)}\left(0,\frac{1}{2};0\right)
    =
    K_z^{(2)}\left(\frac{1}{2},0;0\right)
    &=
    -\frac{\sqrt{z}}{2}
    \operatorname{csch}\left(\frac{\sqrt{z}}{2}\right).
    \label{eq:app_two_source_cross_kernel}
\end{align}
It is useful to introduce the specific combinations
\begin{subequations}
\label{eq:app_auxiliary_functions}
\begin{align}
    \mathcal X(\omega)
    &=
    \alpha
    \coth\left(\frac{\alpha}{2}\right)
    -
    \beta
    \coth\left(\frac{\beta}{2}\right),
    \label{eq:app_def_X}
    \\[1ex]
    \mathcal Y
    &=
    \alpha
    \operatorname{csch}\left(\frac{\alpha}{2}\right),
    \label{eq:app_def_Y}
    \\[1ex]
    \mathcal Z(\omega)
    &=
    \beta
    \operatorname{csch}\left(\frac{\beta}{2}\right).
    \label{eq:app_def_Z}
\end{align}
\end{subequations}
Then the elements of the inverse response matrix can be written compactly as
\begin{subequations}
\label{eq:app_two_source_response_elements}
\begin{align}
    \widetilde S_{11}(\omega)
    &=
    \ii\omega
    +\frac{\lambda_1}{2}
    \left[
        s_1\mathcal X(\omega)
        +s_2\mathcal Y
    \right],
    \label{eq:app_two_source_S11}
    \\[1ex]
    \widetilde S_{12}(\omega)
    &=
    -\frac{\lambda_1s_2}{2}
    \mathcal Z(\omega),
    \label{eq:app_two_source_S12}
    \\[1ex]
    \widetilde S_{21}(\omega)
    &=
    -\frac{\lambda_2s_1}{2}
    \mathcal Z(\omega),
    \label{eq:app_two_source_S21}
    \\[1ex]
    \widetilde S_{22}(\omega)
    &=
    \ii\omega
    +\frac{\lambda_2}{2}
    \left[
        s_2\mathcal X(\omega)
        +s_1\mathcal Y
    \right].
    \label{eq:app_two_source_S22}
\end{align}
\end{subequations}
The determinant of the inverse response matrix is
\begin{equation}
\det\widetilde{\boldsymbol S}(\omega)
=
\widetilde S_{11}(\omega)
\widetilde S_{22}(\omega)
-
\widetilde S_{12}(\omega)
\widetilde S_{21}(\omega).
\label{eq:app_two_source_determinant_definition}
\end{equation}

\begin{multline}
\det\widetilde{\boldsymbol S}(\omega)
=
-\omega^2
+\frac{\ii\omega}{2}
(\lambda_1s_1+\lambda_2s_2)\mathcal X(\omega)
+\frac{\ii\omega}{2}
(\lambda_1s_2+\lambda_2s_1)\mathcal Y
\\
+\frac{\lambda_1\lambda_2s_1s_2}{4}
\left[
    \mathcal X(\omega)^2
    +\mathcal Y^2
    -\mathcal Z(\omega)^2
\right]
+\frac{\lambda_1\lambda_2(s_1^2+s_2^2)}{4}
\mathcal X(\omega)\mathcal Y.
\label{eq:app_two_source_determinant_expanded}
\end{multline}

Comparison with
Eq.~\eqref{eq:two_source_stationary_determinant} identifies the four
frequency-dependent coefficients as follows:
\begin{subequations}
\label{eq:app_response_coefficients}

\begin{equation}
\mathcal A(\omega)
=
\frac{\ii\omega}{2}\mathcal X(\omega).
\label{eq:app_coefficient_A}
\end{equation}

\begin{equation}
\mathcal B(\omega)
=
\frac{\ii\omega}{2}\mathcal Y.
\label{eq:app_coefficient_B}
\end{equation}

\begin{equation}
\mathcal C(\omega)
=
\frac{1}{4}
\left[
    \mathcal X(\omega)^2
    +\mathcal Y^2
    -\mathcal Z(\omega)^2
\right].
\label{eq:app_coefficient_C}
\end{equation}

\begin{equation}
\mathcal D(\omega)
=
\frac{1}{4}
\mathcal X(\omega)\mathcal Y.
\label{eq:app_coefficient_D}
\end{equation}

\end{subequations}

The expression for $\mathcal C(\omega)$ can be simplified using
\begin{equation}
\coth^2x-\operatorname{csch}^2x=1
\end{equation}
together with
\begin{equation}
\beta^2-\alpha^2=\ii\omega.
\end{equation}
From the definitions of $\mathcal Y$ and $\mathcal Z(\omega)$,
\begin{align}
\mathcal Y^2
&=
\alpha^2
\left[
\coth^2\left(\frac{\alpha}{2}\right)-1
\right],
\qquad
\mathcal Z(\omega)^2
=
\beta^2
\left[
\coth^2\left(\frac{\beta}{2}\right)-1
\right].
\end{align}
Combining these expressions with the definition of
$\mathcal X(\omega)$ gives
\begin{align}
\mathcal X(\omega)^2
+\mathcal Y^2
-\mathcal Z(\omega)^2
={}&
\ii\omega
+
2\alpha
\coth\left(\frac{\alpha}{2}\right)
\mathcal X(\omega).
\label{eq:app_C_simplification}
\end{align}
Consequently,
\begin{equation}
\mathcal C(\omega)
=
\frac{\ii\omega}{4}
+
\frac{\alpha}{2}
\coth\left(\frac{\alpha}{2}\right)
\mathcal X(\omega).
\end{equation}
Finally, substituting
\begin{equation}
\alpha=\sqrt{\mu},
\qquad
\beta=\sqrt{\mu+\ii\omega}
\end{equation}
recovers the explicit expressions for
$\mathcal A(\omega)$--$\mathcal D(\omega)$ given in the main text.

\bibliographystyle{apsrev4-2}
\bibliography{references}

@article{activematter,
   author = "Ramaswamy, Sriram",
   title = "The Mechanics and Statistics of Active Matter", 
   journal= "Annu. Rev. Condens. Matter Phys.",
   year = "2010",
   volume = "1",
   pages = "323-345",
   doi = "https://doi.org/10.1146/annurev-conmatphys-070909-104101",
   url = "https://www.annualreviews.org/content/journals/10.1146/annurev-conmatphys-070909-104101",
   publisher = "Annual Reviews",
  }

@article{ActiveColloidsReview,
author = {Illien, Pierre and Golestanian, Ramin and Sen, Ayusman},
year = {2017},
month = {04},
pages = {5508-5518},
title = {‘Fuelled’ motion: phoretic motility and collective behaviour of active colloids},
volume = {46},
journal = {Chem. Soc. Rev.},
doi = {10.1039/c7cs00087a}
}

@article{pimienta,
title = {Self-propulsion on liquid surfaces},
journal = {Curr. Opin. Colloid Interface Sci.},
volume = {19},
number = {4},
pages = {290-299},
year = {2014},
issn = {1359-0294},
doi = {https://doi.org/10.1016/j.cocis.2014.04.001},
url = {https://www.sciencedirect.com/science/article/pii/S1359029414000399},
author = {Pimienta, Véronique and Antoine, Charles},
}

@article{interfacialswimmers,
  title = {Role of Marangoni forces in the velocity of symmetric interfacial swimmers},
  author = {Boniface, Dolachai and Cottin-Bizonne, C{\'e}cile and Detcheverry, Fran{\ifmmode \mbox{\c{c}}\else \c{c}\fi{}}ois and Ybert, Christophe},
  journal = {Phys. Rev. Fluids},
  volume = {6},
  issue = {10},
  pages = {104006},
  numpages = {34},
  year = {2021},
  month = {Oct},
  publisher = {American Physical Society},
  doi = {10.1103/PhysRevFluids.6.104006},
  url = {https://link.aps.org/doi/10.1103/PhysRevFluids.6.104006}
}

@article{reviewphoresis,
  title        = {Phoretic Self-Propulsion},
  journal      = {Annu. Rev. Fluid Mech.},
  year         = {2017},
  volume       = {49},
  author = {Moran, Jeffrey L. and Posner, Jonathan D.},
  pages  = {511--540},
  doi    = {10.1146/annurev-fluid-122414-034456},
  url    = {https://doi.org/10.1146/annurev-fluid-122414-034456},
}

@article{Anderson,
  author       = {J. L. Anderson},
  title        = {Colloid transport by interfacial forces},
  journal      = {Annu. Rev. Fluid Mech.},
  year         = {1989},
  volume       = {21},
  pages        = {61--99},
  url = "https://www.annualreviews.org/content/journals/10.1146/annurev.fl.21.010189.000425",
}

@article{Michelin,
  author       = {S. Michelin and E. Lauga and D. Bartolo},
  title        = {Spontaneous autophoretic motion of isotropic particles},
  journal      = {Phys. Fluids},
  year         = {2013},
  volume       = {25},
  pages        = {061701},
  url={http://dx.doi.org/10.1063/1.4810749},
}

@article{HarmonicTrap,
  title = {Self-phoretic oscillatory motion in a harmonic trap},
  author = {Alexandre, Arthur and Anderson, Leah and Collin-Dufresne, Thomas and Gu\'erin, Thomas and Dean, David S.},
  journal = {Phys. Rev. E},
  volume = {109},
  issue = {6},
  pages = {064147},
  numpages = {9},
  year = {2024},
  month = {Jun},
  publisher = {American Physical Society},
  doi = {10.1103/PhysRevE.109.064147},
  url = {https://link.aps.org/doi/10.1103/PhysRevE.109.064147}
}

@article{1dchannel,
    author = {Anderson, Leah and Dean, David S.},
    title = {Self-phoretic oscillatory motion in a one-dimensional channel},
    journal = {J. Chem. Phys.},
    volume = {165},
    number = {4},
    pages = {044119},
    year = {2026},
    month = {07},
    issn = {0021-9606},
    doi = {10.1063/5.0330797},
    url = {https://doi.org/10.1063/5.0330797},
}

@article{Grima,
  title = {Phase transitions and superuniversality in the dynamics of a self-driven particle},
  author = {Grima, R.},
  journal = {Phys. Rev. E},
  volume = {74},
  issue = {1},
  pages = {011125},
  numpages = {6},
  year = {2006},
  month = {Jul},
  publisher = {American Physical Society},
  doi = {10.1103/PhysRevE.74.011125},
  url = {https://link.aps.org/doi/10.1103/PhysRevE.74.011125}
}

@article{Sengupta,
  title = {Dynamics of a microorganism moving by chemotaxis in its own secretion},
  author = {Sengupta, Ankush and van Teeffelen, Sven and L\"owen, Hartmut},
  journal = {Phys. Rev. E},
  volume = {80},
  issue = {3},
  pages = {031122},
  numpages = {9},
  year = {2009},
  month = {Sep},
  publisher = {American Physical Society},
  doi = {10.1103/PhysRevE.80.031122},
  url = {https://link.aps.org/doi/10.1103/PhysRevE.80.031122}
}

@article{Taktikos,
  title = {Modeling a self-propelled autochemotactic walker},
  author = {Taktikos, Johannes and Zaburdaev, Vasily and Stark, Holger},
  journal = {Phys. Rev. E},
  volume = {84},
  issue = {4},
  pages = {041924},
  numpages = {11},
  year = {2011},
  month = {Oct},
  publisher = {American Physical Society},
  doi = {10.1103/PhysRevE.84.041924},
  url = {https://link.aps.org/doi/10.1103/PhysRevE.84.041924}
}

@article{ChemotaxisSwimmers,
  author       = {C. Jin and C. Krüger and C. C. Maass},
  title        = {Chemotaxis and autochemotaxis of self-propelling droplet swimmers},
  journal      = {Proc. Natl. Acad. Sci.},
  year         = {2017},
  volume       = {114},
  pages        = { 5089-5094},
  url          = {https://doi.org/10.1073/pnas.1619783114},
}

@article{SotoGolestanian2014,
  title = {Self-Assembly of Catalytically Active Colloidal Molecules: Tailoring Activity Through Surface Chemistry},
  author = {Soto, Rodrigo and Golestanian, Ramin},
  journal = {Phys. Rev. Lett.},
  volume = {112},
  issue = {6},
  pages = {068301},
  numpages = {5},
  year = {2014},
  month = {Feb},
  publisher = {American Physical Society},
  doi = {10.1103/PhysRevLett.112.068301},
  url = {https://link.aps.org/doi/10.1103/PhysRevLett.112.068301}
}

@incollection{Liebchen2019,
  author    = {Liebchen, Benno and L{\"o}wen, Hartmut},
  title     = {Modeling Chemotaxis of Microswimmers: From Individual to Collective Behavior},
  booktitle = {Chemical Kinetics: Beyond the Textbook},
  publisher = {World Scientific Publishing Europe Ltd.},
  address   = {London},
  year      = {2019},
  pages     = {493--516},
  doi       = {10.1142/9781786347015_0020},
  url       = {https://doi.org/10.1142/9781786347015_0020}
}

@article{Kohira2001,
  author  = {Kohira, M. I. and Hayashima, Y. and Nagayama, M. and Nakata, S.},
  title   = {Synchronized Self-Motion of Two Camphor Boats},
  journal = {Langmuir},
  volume  = {17},
  number  = {23},
  pages   = {7124--7129},
  year    = {2001},
  doi     = {10.1021/la010388r}
}

@Article{Camphorboat1,
    AUTHOR = {Roliński, Tomasz and Kitahata, Hiroyuki and Koyano, Yuki and Górecki, Jerzy},
    TITLE = {Quantitative Analysis of the Complex Time Evolution of a Camphor Boat},
    VOLUME = {14},
    YEAR = {2024},
    NUMBER = {3},
    URL = {https://www.mdpi.com/2076-3417/14/3/959},
    journal = {Appl. Sci.},
    pages   = {959},
    doi     = {10.3390/app14030959},
}

@article{Camphorboat2,
  author       = { Nakata, Satoshi and Iguchi, Yasutaka and Ose, Sachie and Kuboyama, Makiko and Ishii, Toshio and Yoshikawa, Kenichi },
  title        = {Self-rotation of a camphor scraping on water: New insight into the old problem},
  journal      = {Langmuir},
  year         = {1997},
  volume       = {13},
  pages = {4454--4458},
    url ={https://doi.org/10.1021/LA970196P},
}

@article{Suematsu2010Benzoquinone,
  author  = {Suematsu, N. J. and Miyahara, Y. and Matsuda, Y. and Nakata, S.},
  title   = {Self-Motion of a Benzoquinone Disk Coupled with a Redox Reaction},
  journal = {J. Phys. Chem. C},
  volume  = {114},
  number  = {31},
  pages   = {13340--13343},
  year    = {2010},
  doi     = {10.1021/jp104666b}
}

@article{Iida2010Phenanthroline,
  author  = {Iida, K. and Suematsu, N. J. and Miyahara, Y. and
             Kitahata, H. and Nagayama, M. and Nakata, S.},
  title   = {Experimental and Theoretical Studies on the Self-Motion of a
             Phenanthroline Disk Coupled with Complex Formation},
  journal = {Phys. Chem. Chem. Phys.},
  volume  = {12},
  number  = {7},
  pages   = {1557--1563},
  year    = {2010},
  doi     = {10.1039/B918691C}
}

@article{runandtumble,
  author  = {Tailleur, J. and Cates, M. E.},
  title   = {Statistical Mechanics of Interacting Run-and-Tumble Bacteria},
  journal = {Phys. Rev. Lett.},
  volume  = {100},
  number  = {21},
  pages   = {218103},
  year    = {2008},
  doi     = {10.1103/PhysRevLett.100.218103}
}

@article{activebrownian,
  author  = {ten Hagen, B. and van Teeffelen, S. and L{\"o}wen, H.},
  title   = {Brownian Motion of a Self-Propelled Particle},
  journal = {J. Phys.: Condens. Matter},
  volume  = {23},
  number  = {19},
  pages   = {194119},
  year    = {2011},
  doi     = {10.1088/0953-8984/23/19/194119}
}

@article{DiskDetcheverry,
  title = {Self-propulsion of symmetric chemically active particles: Point-source model and experiments on camphor disks},
  author = {Boniface, Dolachai and Cottin-Bizonne, C\'ecile and Kervil, Ronan and Ybert, Christophe and Detcheverry, Francois},
  journal = {Phys. Rev. E},
  volume = {99},
  issue = {6},
  pages = {062605},
  numpages = {13},
  year = {2019},
  month = {Jun},
  publisher = {American Physical Society},
  doi = {10.1103/PhysRevE.99.062605},
  url = {https://link.aps.org/doi/10.1103/PhysRevE.99.062605}
}

@article{Koyano1,
  title = {Oscillatory motion of a camphor grain in a one-dimensional finite region},
  author = {Koyano, Yuki and Sakurai, Tatsunari and Kitahata, Hiroyuki},
  journal = {Phys. Rev. E},
  volume = {94},
  issue = {4},
  pages = {042215},
  numpages = {8},
  year = {2016},
  month = {Oct},
  publisher = {American Physical Society},
  doi = {10.1103/PhysRevE.94.042215},
  url = {https://link.aps.org/doi/10.1103/PhysRevE.94.042215}
}

@article{CamphorGrain1dbox,
  title = {A Camphor Grain Oscillates while Breaking Symmetry},
  author = {Y. Hayashima and M. Nagayama and S. Nakata},
  journal = {J. Phys. Chem. B},
  volume = {105},
  pages = {5353-5357},
  year = {2001},
  url  = {https://doi.org/10.1021/jp004505n},
}

@article{Nagayama2004,
  author  = {Nagayama, M. and Nakata, S. and Doi, Y. and Hayashima, Y.},
  title   = {A Theoretical and Experimental Study on the Unidirectional Motion of a Camphor Disk},
  journal = {Physica D},
  volume  = {194},
  number  = {3--4},
  pages   = {151--165},
  year    = {2004},
  doi     = {10.1016/j.physd.2004.02.003}
}

@article{Frank2014,
  author  = {Frank, T. D.},
  title   = {Exact Solutions for Chemical Concentration Waves of Self-Propelling Camphor Particles Racing on a Ring: A Novel Potential Dynamics Perspective},
  journal = {Condens. Matter Phys.},
  volume  = {17},
  number  = {4},
  pages   = {43002},
  year    = {2014},
  doi     = {10.5488/CMP.17.43002}
}

@article{Nishi2015,
  author  = {Nishi, K. and Wakai, K. and Ueda, T. and Yoshii, M. and
             Ikura, Y. S. and Nishimori, H. and Nakata, S. and Nagayama, M.},
  title   = {Bifurcation Phenomena of Two Self-Propelled Camphor Disks on an Annular Field Depending on System Length},
  journal = {Phys. Rev. E},
  volume  = {92},
  number  = {2},
  pages   = {022910},
  year    = {2015},
  doi     = {10.1103/PhysRevE.92.022910}
}

@article{Yasugahira2022,
  author  = {Yasugahira, Y. and Nagayama, M.},
  title   = {On a Numerical Bifurcation Analysis of a Particle Reaction-Diffusion Model for a Motion of Two Self-Propelled Disks},
  journal = {Jpn. J. Ind. Appl. Math.},
  volume  = {39},
  pages   = {631--652},
  year    = {2022},
  doi     = {10.1007/s13160-021-00498-4}
}

@article{Suematsu2010,
  author  = {Suematsu, N. J. and Nakata, S. and Awazu, A. and Nishimori, H.},
  title   = {Collective Behavior of Inanimate Boats},
  journal = {Phys. Rev. E},
  volume  = {81},
  number  = {5},
  pages   = {056210},
  year    = {2010},
  doi     = {10.1103/PhysRevE.81.056210}
}

@article{Heisler2012,
  author  = {Heisler, E. and Suematsu, N. J. and Awazu, A. and Nishimori, H.},
  title   = {Swarming of Self-Propelled Camphor Boats},
  journal = {Phys. Rev. E},
  volume  = {85},
  number  = {5},
  pages   = {055201},
  year    = {2012},
  doi     = {10.1103/PhysRevE.85.055201}
}

@article{Ikura2013,
  author  = {Ikura, Y. S. and Heisler, E. and Awazu, A. and
             Nishimori, H. and Nakata, S.},
  title   = {Collective Motion of Symmetric Camphor Papers in an Annular Water Channel},
  journal = {Phys. Rev. E},
  volume  = {88},
  number  = {1},
  pages   = {012911},
  year    = {2013},
  doi     = {10.1103/PhysRevE.88.012911}
}

@article{Ikeda2019,
  author  = {Ikeda, K. and Ei, S.-I. and Nagayama, M. and
             Okamoto, M. and Tomoeda, A.},
  title   = {Reduced Model of a Reaction-Diffusion System for the Collective Motion of Camphor Boats},
  journal = {Phys. Rev. E},
  volume  = {99},
  number  = {6},
  pages   = {062208},
  year    = {2019},
  doi     = {10.1103/PhysRevE.99.062208}
}

\end{document}